# Strongly anisotropic strain-tunability of excitons in exfoliated ZrSe₃


*Hao Li, Gabriel Sanchez-Santolino, Sergio Puebla, Riccardo Frisenda, Abdullah M. Al-Enizi, Ayman Nafady, Roberto D'Agosta*, Andres Castellanos-Gomez**

Hao Li, Sergio Puebla, Dr. Riccardo Frisenda, Dr. Andres Castellanos-Gomez
Materials Science Factory. Instituto de Ciencia de Materiales de Madrid (ICMM-CSIC),
Madrid, E-28049, Spain.
E-mail: Andres.castellanos@csic.es

Gabriel Sanchez-Santolino
GFMC, Departamento de Física de Materiales & Instituto Pluridisciplinar, Universidad
Complutense de Madrid, 28040 Madrid, Spain

Abdullah M. Al-Enizi, Ayman Nafady
Department of Chemistry, College of Science, King Saud University, Riyadh 11451, Saudi
Arabia

Prof. Roberto D'Agosta
Nano-Bio Spectroscopy Group and European Theoretical Spectroscopy Facility (ETSF),
Departamento de Polímeros y Materiales Avanzados: Física, Química y Tecnología,
Universidad del País Vasco UPV/EHU, Avenida Tolosa 72, E-20018 San Sebastián, Spain
IKERBASQUE, Basque Foundation for Science, Plaza Euskadi 5, E-48009 Bilbao, Spain
E-mail: roberto.dagosta@ehu.es





We study the effect of uniaxial strain on the band structure of ZrSe₃, a semiconducting material with a marked in-plane structural anisotropy. By using a modified 3-point bending test apparatus, thin ZrSe₃ flakes were subjected to uniaxial strain along different crystalline orientations monitoring the effect of strain on their optical properties through micro-reflectance spectroscopy. The obtained spectra showed excitonic features that blueshift upon uniaxial tension. This shift is strongly dependent on the direction along which the strain is being applied. When the flakes are strained along the b-axis, the exciton peak shifts at ~ 60-95 meV/%, while along the a-axis, the shift only reaches ~ 0-15 meV/%. Ab initio calculations were conducted to study the influence of uniaxial strain, applied along different crystal directions, on the band structure and reflectance spectra of ZrSe₃, exhibiting a remarkable agreement with the experimental results.


## 1. Introduction

Applying mechanical deformations has become a powerful approach to modify the vibrational, optical, and electronic properties of two-dimensional (2D) materials.[1–7] In principle, 2D materials can sustain unprecedented strains without breaking.[8–10] Moreover, their band





structures are rather strain-sensitive,[11–14] making them very suitable for strain engineering applications. In addition, 2D materials offer the possibility to apply strain in various ways (e.g., uniaxial/biaxial, homogenous/inhomogeneous), which can be adjusted by the end-user at will. Strain engineering in conventional 3D materials, in contrast, typically relies on forcing the epitaxial growth of a material onto a substrate with a given lattice parameter mismatch, providing a fixed strain that cannot be adjusted after growth. Among the different strategies to strain engineer 2D materials, the application of uniaxial strain through bending flexible substrates with a bending jig apparatus is one of the most popular approaches.[11,15–20] This method presents some issues when applied to 2D materials with in-plane anisotropic properties. Indeed, the effect of uniaxial strain along different crystal orientations is expected to modify the properties of these anisotropic 2Ds differently. Despite the recent interest on these families of anisotropic 2D materials,[21–28] the number of reported research works focused on studying the effect of strain along different crystal directions is still very scarce and primarily focused on the investigation of strain tunable Raman modes in black phosphorus, PdSe$_2$ or tellurium.[29–33] The group IV–V transition metal trichalcogenides (TMTC) is a less-explored family of materials with quasi-1D electrical and optical properties stemming from a reduced in-plane structural symmetry.[26,34–36] These materials have a general formula of MX$_3$ being M a transition metal atom belonging to either group IVB (Ti, Zr, Hf) or group VB (Nb, Ta) and X chalcogen atoms from group VIA (S, Se, Te).

In this work, we focused on ZrSe$_3$, a semiconductor of the transition metal tri-chalcogenide family[26] with a strong in-plane anisotropic structure similar to that of TiS$_3$,[37–40] along different crystal directions (see Figure 1a). ZrSe$_3$ has been scarcely studied experimentally so far. Xiong et al. presented a photodetector based on individual ZrSe$_3$ nanobelts[41] and Zhou et al. proposed its potential application in thermoelectrics through calculations,[42] but little is still known about





the fundamental electronic and optical properties of this semiconducting material. We explored how applying strain along different crystal orientations modifies the excitonic features in the optical spectra. Our straining technique relies on a three-point bending test apparatus, and it can be easily adapted by other groups that are already working on strain engineering of 2D materials. Using this straining approach, we demonstrate how the direction along which we apply the uniaxial strain has a strong effect on the strain tunability of the excitonic features on this anisotropic 2D material. Evidently, a giant anisotropy in the strain-induced exciton shift was found. The strain gauge factor varies from ~ 60-95 meV/% for uniaxial strain applied along the *b*-axis to ~ 0-15 meV/% when the uniaxial strain is applied along the *a*-axis. Additionally, *ab-initio* calculations were performed, supporting the finding of the strongly anisotropic strain-tunable direct band-gap transitions observed in the experiments. Therefore, being able to accurately align the direction of the uniaxial strain axis with specific crystalline orientations is strongly appealing to further employ this family of 2D materials in strain engineering applications.

## 2. Sample fabrication and straining setup

$ZrSe_3$ flakes are obtained by mechanical exfoliation of bulk crystals and transferred through an all-dry deterministic placement method[43–45] (see Materials and Methods section). Figure 1b shows a high-angle annular dark-field scanning transmission electron microscopy (HAADF-STEM) image of a thin $ZrSe_3$ flake transferred onto a holey $Si_3N_4$ TEM grid to characterize its crystal structure. Figure 1c shows an atomic resolution HAADF image of the same flake along the (001) direction, in which the anisotropic in-plane structure of $ZrSe_3$ is visible. The inset shows a selected area diffraction pattern (SAED) acquired at the same region corresponding to a monoclinic ($P2_1/m$) crystal structure. A high magnification atomic resolution HAADF image





shown in Figure 1d clearly illustrates the chain-like structure of ZrSe$_3$ within the *a-b* plane, similar to that of TiS$_3$.[26]

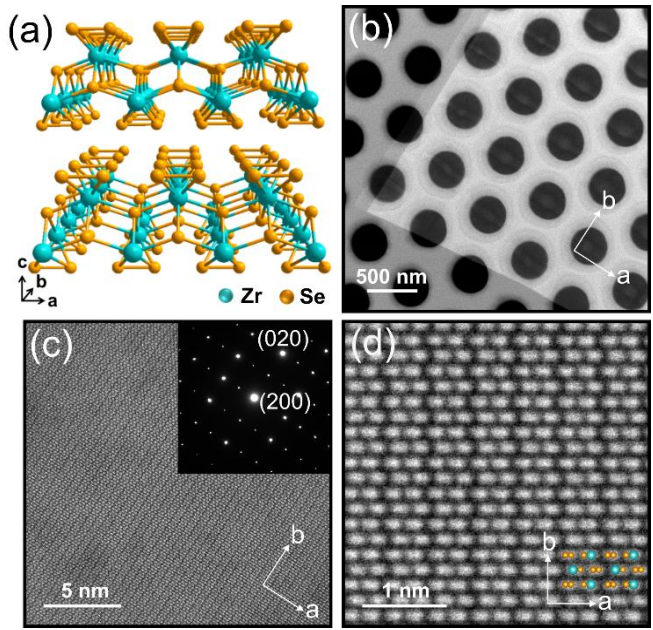

**Figure 1. Structural characterization of ZrSe$_3$.** (a) 3D representation of the crystal structure of ZrSe$_3$ where the in-plane anisotropy can be resolved. (b, c) Low magnification and atomic resolution HAADF images of a mechanically exfoliated ZrSe$_3$ flake transferred over a holey Si$_3$N$_4$ membrane. The inset in (c) shows a SAED pattern acquired at the same region; (d) Atomic resolution HAADF image of the ZrSe$_3$ flake down the (001) direction depicting the chains-like structure along the *b*-axis. The superimposed atomic model depicts the Zr (cyan) and Se (orange) atomic column positions.

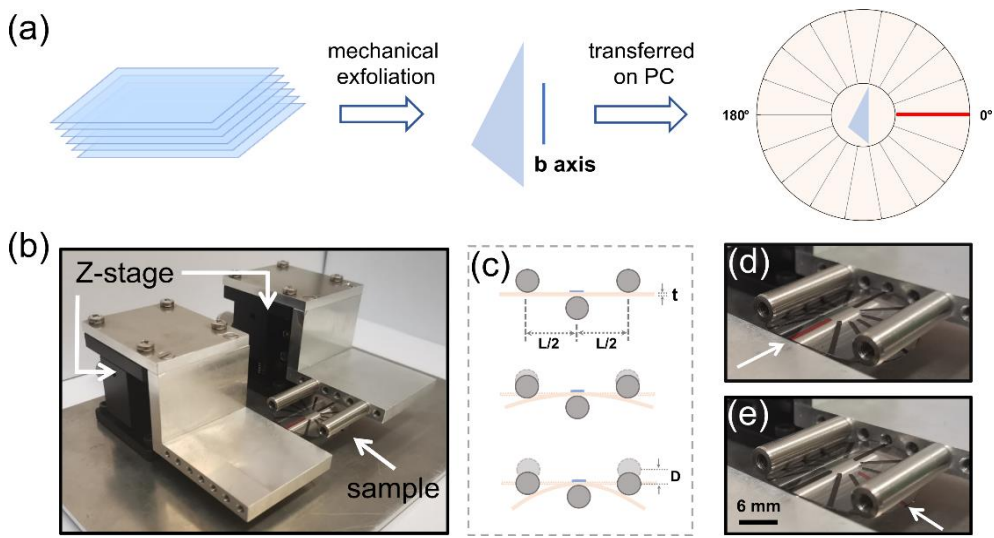

**Figure 2. Setup for angle-resolved uniaxial straining measurements.** (a) Cartoon of the fabrication process of the test sample. A disc-shaped polycarbonate substrate is marked with a permanent marker every 20º angle from 0º-360º, highlighting the 0º with a red line. An exfoliated ZrSe$_3$ flake is then transferred onto the center of the disc by an all-dry deterministic placement method. (b) Picture of the three-points bending setup used for the strain engineering experiments with a disc-shaped flexible substrate loaded between the pivotal points. (c) Schematic





representation of the three-point bending process. (d, e) Pictures of the sample upon 0.65% of uniaxial strain applied along two orthogonal directions (notice the position of the red marker line).

The angle-dependent uniaxial strain measurements are based on the integration of disc-shaped flexible substrates, instead of the commonly used rectangular beam, in a bending-test apparatus. Figure 2a is a schematic representation of the fabrication process. The samples are fabricated out of a disc-shaped polycarbonate substrate (thickness = 250 μm). Permanent marker lines are drawn on the surface to guide the angle adjustment. The ZrSe$_3$ flakes are then transferred onto the geometrical center of the disc surface by means of an all-dry deterministic placement method with an accuracy of ~10 μm.[43–45]

The disc-shaped substrate with the ZrSe$_3$ is then loaded in a homebuilt three-point bending jig apparatus to apply uniaxial strain through bending of the substrate (Figures 2b and 2c).[16] After a strain load/unload cycle, the disc-shaped sample can be rotated to apply uniaxial strain along another direction. Figure 2d and 2e show an example where uniaxial strain is applied along two orthogonal directions (notice the position of the red marker line). The amount of strain can be extracted from the geometry of the disc substrate and the bending apparatus through:

$$\varepsilon = \frac{6tD}{L^2},$$

where $t$ is the thickness of the substrate, $L$ the distance between the outer pivotal points and $D$ the deflection of the central pivotal point (see Figure 2c). Note that we have experimentally validated this formula to calculate the strain in Ref. [16] by directly measuring the separation of micro-fabricated features in the surface of the polycarbonate substrates upon controlled deflection of the substrate with the 3-point bending apparatus.

**3. Results**





Figure 3a displays an optical microscopy image of a ~16 nm thick ZrSe₃ flake (see Supporting Information Figure S1), transferred onto the center of a disc-shaped flexible polycarbonate substrate. For the angle-dependent uniaxial straining experiments, it is crucial to accurately determine the crystal orientation of the ZrSe₃ flake under study. This can be done by measuring reflection or absorption spectra with linearly polarized incident light. When the incident light is linearly polarized along the *b*-axis the spectra show prominent excitonic features, even at room temperature.[46,47] When the light is polarized along the *a*-axis, on the other hand, the intensity of these excitonic features decreases substantially.

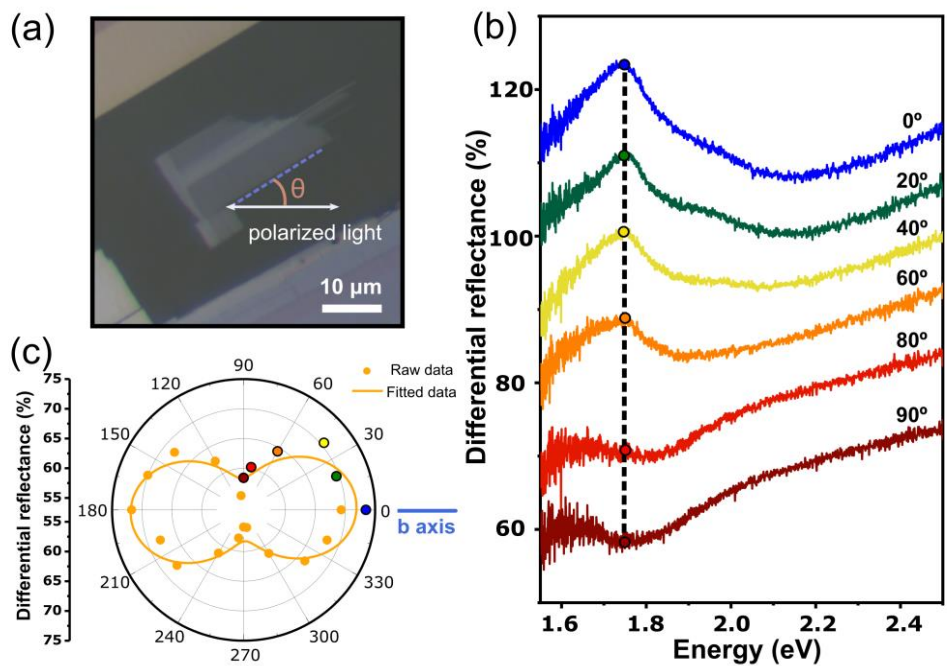

**Figure 3. Identification of crystalline orientations of ZrSe₃.** (a) Optical microscopy image of a few-layer ZrSe₃ flake. In the angle-resolved measurements, θ is defined as the angle between the cleaved long straight edge of the flake (dashed blue line) and the polarization direction (parallel to the horizontal axis of the microscope, white arrow). (b) Micro-reflectance spectra of the ZrSe₃ flake (unstrained) as a function of sample rotation angle from 0° to 90° while the incident light is linearly polarized parallel to the horizontal axis. The spectra have been vertically offset by 10% to facilitate the comparison. (c) Polar plot of the differential reflectance intensity at ~1.75 eV for different angles between the incident linearly polarized light and the cleaved long straight edge of the ZrSe₃ flake. The color circles correspond to the spectra shown in (b).





We have used a homebuilt differential reflectance system to acquire reflectance spectra with linearly polarized incident light to determine the $ZrSe_3$ crystal orientation.[48,49] We found that other optical spectroscopic techniques, like photoluminescence or Raman spectroscopy, could not be carried out in the $ZrSe_3$ due to laser-induced damage of the flakes even at very low laser power density values (~ 40 $\mu W/\mu m^2$). Supporting Information Figure S2 shows the damage caused by the laser when attempting to carry out Raman spectroscopy and photoluminescence measurements with low excitation power. In our differential reflectance experiments we fixed the incident light linearly polarized parallel to the horizontal axis of the microscope while the $ZrSe_3$ flake is rotated to vary the orientation between the linearly polarized light and the crystal axes. Figure 3b shows differential reflectance spectra acquired for different angles ($\theta$) between the incident linearly polarized light and the long cleaved edge of the $ZrSe_3$ flake (highlighted with a dashed blue line in Figure 3a). When the straight long edge of the flake is parallel to the linearly polarized light (labelled as 0º or 180º), the spectra show a broad peak feature at ~1.75 eV and a shoulder at ~1.9 eV. These peaks are attributed to the generation of excitons (called A and B in the literature, respectively) due to direct valence-to-conduction band transitions.[46,47] The intensity of these excitonic peaks decreases when the flake is rotated and it reaches its minimum intensity when the long edge of the flake is perpendicular to the incident linearly polarized light. This observation indicates that the *b*-axis of the $ZrSe_3$ flake is oriented parallel to the long edge of the flake.[46,47] Figure 3c represents the intensity of the A exciton peak at different angles between the incident linearly polarized light and the long straight edge of the flake. The polar plot clearly illustrates how the intensity of the exciton feature reaches its maximum value when the long edge of the flake is aligned parallel to the incident polarized light (horizontal axis).





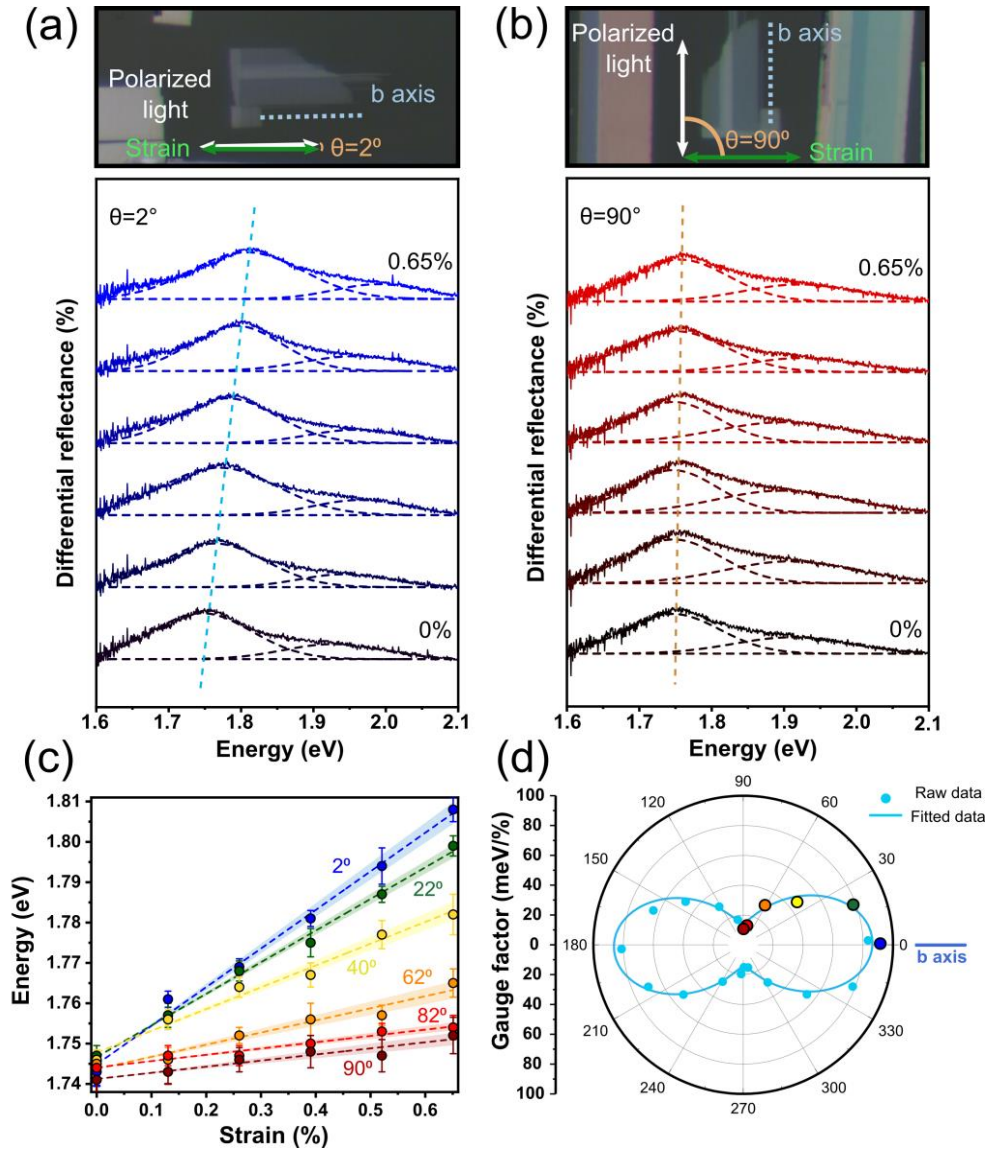

**Figure 4. Angle-resolved micro-reflectance spectra of ZrSe₃ under different uniaxial strain from 0%-0.65%.** (a, b) Micro-reflectance spectra acquired when the uniaxial strain direction is parallel to the *b*-axis and the *a*-axis of the ZrSe₃ flake, respectively. (c) A-exciton peak energy as a function of the applied uniaxial strain under different orientation angles (2º, 22º, 40º, 62º, 82º and 90º). A linear fit is used to extract the gauge factor. The shaded area around the dashed lines indicates the uncertainty in the linear fit and provides a measurement of the uncertainty of the extracted gauge factor. (d) Angular dependence of the ZrSe₃ A exciton gauge factor in polar coordinates. All the tests are carried out with the strain axis parallel to the horizontal axis of the microscope and by keeping the incident linearly polarized light and flake *b*-axis direction parallel to each other along the whole experiment to maximize the excitons intensities.

After determining the crystal orientations of the ZrSe₃ flake under study, we subject it to uniaxial strain cycles. Differential reflectance spectroscopy at different strain values was conducted to infer the effect of strain on the excitonic features of the optical spectra. After each uniaxial strain cycle, the disc-shaped substrate is rotated by ~20º, changing the angle (θ)





between the crystal *b*-axis and the uniaxial strain direction. We label as 0º/180º the situation

where the uniaxial strain is parallel to the *b*-axis and 90º/270º the one where the uniaxial strain

is perpendicular to the *b*-axis. For this experiment, the incident polarized light is kept parallel

to the *b*-axis to maximize the intensity of the exciton peaks but comparable results are obtained

using unpolarized light (see Supporting Information Figure S4).

Figure 4a shows a selection of differential reflectance spectra acquired while applying

different uniaxial strain values almost parallel to the *b*-axis ($\theta = 2º$). Figure 4b shows other

differential reflectance spectra *vs* strain acquired when the uniaxial strain is orthogonal to the

*b*-axis ($\theta = 90º$). Note that a smooth polynomial background has been subtracted from the

spectra to facilitate identifying the exciton peaks (see the Supporting Information Figure S3 that

compares some spectra before and after removing the background). The spectra have been fitted

to a sum of two Gaussian peaks to accurately determine the energy of the A and B excitonic

features. Figure 4c summarizes the A exciton peak energy *vs* uniaxial strain acquired for

different alignment between the uniaxial straining direction and the ZrSe$_3$ *b*-axis. The strain

gauge factor, defined as the spectral shift per % of uniaxial strain, strongly depends on the angle

between the uniaxial straining direction and the *b*-axis. It ranges from ~95 meV/% for parallel

configuration ($\theta = 2º$) to ~15 meV/% for the orthogonal configuration ($\theta = 90º$). Figure 4d

further summarizes the gauge factors determined for different alignment between the uniaxial

strain direction and the ZrSe$_3$ *b*-axis.

The B exciton peak is much lower in intensity and it is wider, making the Gaussian fit less

accurate. This motivated us to focus on the A exciton, that contains information about the lower

energy direct band-to-band transition. Nonetheless, we address the reader to Supporting

Information Figure S5 for a dataset where both A and B extion energies have been determined

as a function of strain. Figure S5 also shows the effect of compressive uniaxial strain (using a





four-points bending setup, see Scheme S1) on the differential reflectance of a $ZrSe_3$ flake, showing how the A and B exciton peaks redshift upon compression at a rate of ~ -80 meV/%.

It is worth noting that the $ZrSe_3$ gauge factor for uniaxial strain along the *b*-axis is substantially larger than that of transition metal dichalcogenides (typically in the 30-60 meV/%) [16] and reproducible (see Supporting Information Figure S6). In fact, this value is very close to that of InSe [50,51] and black phosphorus[31], the 2D materials with the largest strain gauge factors reported so far. However, both InSe and black phosphorus are sensitive to the environmental exposure and tend to degrade. We have found that the observed exciton strain tunability remain stable even after 4 months of exposure to atmospheric conditions (see Supporting Information Figures S7 to S10).

We have measured a total of 11 $ZrSe_3$ flakes obtaining comparable results (See Figures S11 to S19 of the Supporting Information). This illustrates the robustness of the observed anisotropy in the strain-tunable A exciton energy. In order to get an insight about the maximum strain that the $ZrSe_3$ flakes can sustain Flakes 7, 8 and 9 were subjected to uniaxial strain until fracture (see Figures S16, S17 and S18 in the Supporting Information), revealing that the flake breakdown occurs between ~ 0.8-1.2%.

To quantify the observed anisotropy in the strain tunability of the excitons we define the anisotropy ratio of the gauge factor as:

$$\text{Anisotropy ratio} = (GF_{max} - GF_{min})/(GF_{max} + GF_{min}) \cdot 100\%,$$

where $GF_{min}$ and $GF_{max}$ stands for the minimum and maximum gauge factors, respectively. This formula gives 0% for perfectly isotropic material. For the studied $ZrSe_3$ flakes, we get values ranging from 72% to ~100%. Black phosphorus, another 2D material with a remarkable in-plane structural anisotropy, showed only a ~3% anisotropy ratio of the gauge factor.[31] Table 1 compares the gauge factors along different crystal directions for $ZrSe_3$ and those reported for





black phosphorus in reference [31]. Table 1 also includes information about the thickness of the ZrSe₃ flakes. Samples 1, 2, 6, 8 and 11 were characterized by AFM. For the other ZrSe₃ samples an estimation of their thickness has been obtained from the optical contrast extracted from the red channel of the optical images (see Supporting Information Figure S20 for details). We have not found any clear thickness dependence in the resulting gauge factor nor anisotropy ratio. We attribute to the fact that the thinner studied flake was ~ 8 nm (~8 layers) that is expected to show bulk-like properties (see Supporting Information S21). At the present we could not exfoliate thinner flakes with a large enough area to study strain engineering experiments with differential reflectance. Note that for a good strain transfer flakes of at least 10×10 μm are selected.

| Reference | Material | Thickness (nm) | Gauge factor (meV/%) | | Anisotropy ratio (%) |
|---|---|---|---|---|---|
| | | | Axis 1 | Axis 2 | |
| This work | ZrSe₃ (Sample 1, Fig. 3) | 16.0 AFM | + (15.1±2.4) | + (95.7±4.1) | 72.7 |
| | ZrSe₃ (Sample 2, Fig. S6) | 14.5 AFM | ~ 0 | + (75.7±5.3) | ~ 100 |
| | ZrSe₃ (Sample 3, Fig. S7) | ~ 8 OPT | + (5.7±3.2) | + (70.2±0.6) | 84.9 |
| | ZrSe₃ (Sample 4, Fig. S8) | ~ 16 OPT | + (4.4±3.9) | + (91.7±8.9) | 90.8 |
| | ZrSe₃ (Sample 5, Fig. S9) | ~ 16 OPT | + (2.0±2.8) | + (82.7±3.9) | 95.3 |
| | ZrSe₃ (Sample 6, Fig. S10) | 8.3 AFM | + (4.6±3.0) | + (92.0±5.7) | 90.5 |
| | ZrSe₃ (Sample 7, Fig. S11) | ~16 OPT | — | + (70.1±4.0) | — |
| | ZrSe₃ (Sample 8, Fig. S12) | 12.8 AFM | — | + (64.1±3.4) | — |
| | ZrSe₃ (Sample 8, Fig. S12) * | 12.8 AFM | + (10.5±4.1) | + (77.6±3.6) | 76.2 |
| | ZrSe₃ (Sample 9, Fig. S13) | ~ 13 OPT | — | + (74.3±5.0) | — |
| | ZrSe₃ (Sample 10, Fig. S14) ** | ~ 15 OPT | — | + (68.7±7.9) to + (81.2±4.7) | — |
| | ZrSe₃ (Sample 10, Fig. S5) *** | ~ 15 OPT | — | -(78.3±4.0) | — |
| | ZrSe₃ (Sample 11, Fig. S5) **** | 14.2 AFM | + (9.6±5.0) to + (22.6±5.0) | + (83.8±2.0) to + (93.0±5.5) | 57.5 to 81.3 |
| | ZrSe₃ (ab initio) | bulk | + 10.9 | + 58.7 | 68.7 |
| [31] | BP | 3.3 | +117 | +124 | 2.9 |

**Table 1.** Summary of the reported exciton shift upon uniaxial strain along different in-plane directions for ZrSe₃ and black phosphorus. *Measurement carried out 4 months after the sample fabrication (sample stored in air). **Measurement acquired in 5 straining/releasing cycles. ***Measurement acquired by applying compressive strain. ****Measurements carried out at different times along 1 month of exposure to air.

We have also carried out *ab-initio* calculations to investigate further this strongly anisotropic response to strain of ZrSe₃. Figure 5a shows the electronic band structure of bulk ZrSe₃ as calculated within the GW approximation for the unstrained structure and for 1% strain applied





along $a$ and $b$ axes. For the unstrained structure, we obtain a direct electronic band-gap of 1.29 eV at the $\Gamma$ point, and an indirect band gap of 0.66 eV between $\Gamma$ and X points. The GW approximation corrects the Density Functional Theory band-gaps (direct, 0.46 eV, indirect 0.15 eV) bringing them closer to the experimental values. The main effect of the GW approximation is a rigid upwards shift the conduction bands in the whole Brillouin zone. We clearly see that a 1% tensile strain along the $b$-axis shifts the direct band gap up by about +90 meV while the same strain along the $a$-axis only shifts the bands by +45 meV. Interestingly, the anisotropy in the strain-tunable indirect band-gap is less pronounced, +100 meV/% along $b$-axis and +85 meV/% along $a$-axis, in good agreement with previously reported DFT calculations that only focused on the indirect band-gap strain tunability without considering excitonic effects.[52]

In order to directly compare with our experimental findings, we have calculated reflectance spectra (Figure 5b and 5c) via solving the Bethe-Salpeter's equation (see Materials and Methods for further information). The spectra present two prominent peaks around ~1.4 eV and ~1.75 eV, due to the generation of A and B excitons respectively. These two peaks are lower in energy than the experimental values, but their positions are consistent with the smaller band-gap we obtained in the GW approximation. Interestingly, our calculation also predicts the presence of another exciton, that we label as B*, with an energy between that of A and B excitons but with a very low intensity. Figure 5a includes a zoomed in plot of the band structure, indicating the different band-to-band transitions that originate the A, B and B* excitons.

We calculated the effect of tensile strain along $a$- and $b$-axis, considering incident light polarized along the $b$-axis of ZrSe$_3$. For the strain along the $b$-axis, we clearly observe a blueshift of the exciton peaks (at a rate of ~59 meV/% of uniaxial strain for the A exciton), consistent with an increase in the electronic band gap. The exciton blueshift is much less pronounced (only ~11 meV/% for the A exciton) when uniaxial strain is applied along the $a$-





axis, reflecting the minimal effect of strain applied along that crystal direction on the band structure. In order to get an insight about the physical origin of the anisotropy in the strain-tunable exciton energies we have analysed the contribution of the different atoms, and their orbitals, to the band structure (see Supporting Information Figures S22 to S24 and related discussion). Our analysis allows us to conclude that the conduction bands, around Γ, are mainly determined by the Zr $d_{xy}$ and $d_{yz}$ orbitals. This fact suggests that deforming the unit cell along y (*b*-axis) might have a more significant impact since it modifies the bonds containing the $d_{xy}$ and $d_{yz}$ orbitals (see Figure S25 in the Supporting Information). At the same time, a deformation along x (*a*-axis) only affects the $d_{xz}$ orbital. Regarding the valence band maximun around the Γ point, our analysis concludes that it is mainly determined by the $p_z$ orbital of $Se_2$ atoms. This observation explains the rather insensitiveness of the valence band maximum to strain applied to both *a* and *b* direction (parallel to x and y in our case) as pointed out by our band structure calculations

Note that the theoretical anisotropy ratio is ~69%, slightly lower than the experimental values that span from 72% to nearly 100%. We attribute this underestimation of the anisotropy to the fact that we neglected the Poisson's effect in our calculation as we could not find any previously peer-reviewed reported value of the Poisson's ratio of $ZrSe_3$. According to the Poisson's effect, a tensile strain along the *a*-axis will be accompanied by a small compression along the *b*-axis that could effectively reduce the exciton strain-induced shift. In fact, a compression along *b*-axis will lead to a redshift of the excitons (see Figure S5 of the Supporting Information) which could partially compensate the blueshift produced by the tension along *a*-axis, thus increasing the anisotropy ratio.





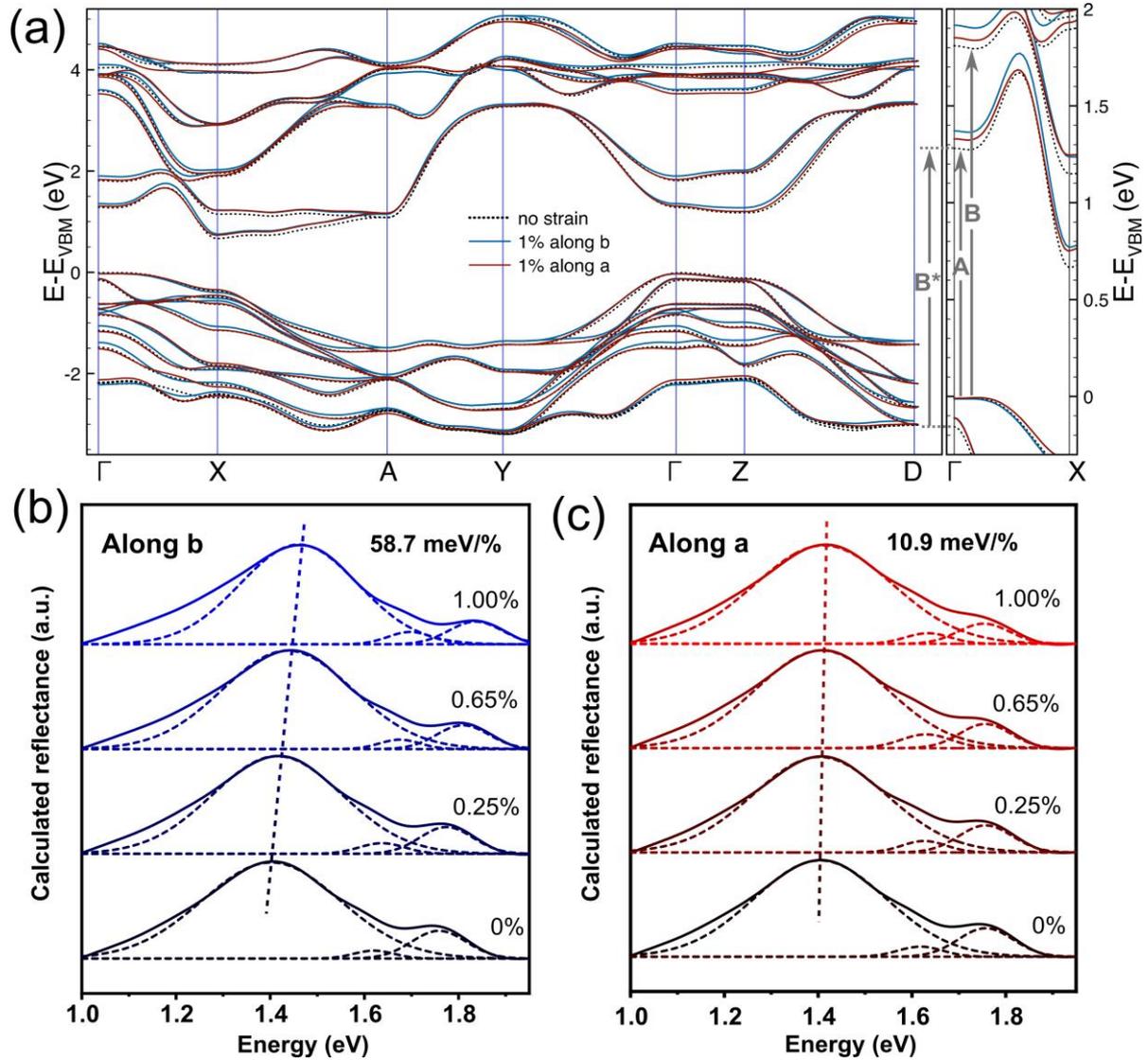

**Figure 5. Calculated band structure and reflectance spectra for uniaxial strains along the *b*- and *a*-axis.** (a) Band structure, calculated within the GW approximation, for different straining situations: 0% strain (dotted black), +1% uniaxial strain along *a*-axis (red) and +1% uniaxial strain along *b*-axis (blue). We have selected the path in the first Brillouin zone: Γ=(0,0,0), X=(½,0,0), A=(½,½,0), Y=(0,½,0), Z=(0,0,½), D=(0,½,½). A zoomed in region in the Γ-X has been plotted to highlight the direct band-to-band transitions that originate the A, B and B* excitons. (b) and (c) Calculated reflectance spectra (solid lines) for different uniaxial tensile strains along the *b*-axis (blue) and *a*-axis (red) respectively. The dashed lines represent a fit with three gaussian functions to determine the position of the A, B and B* exciton peaks.

## 4. Conclusions

In summary, we subjected thin ZrSe$_3$ flakes to uniaxial strain along different crystal orientations. We found that the reflectance spectra showed a peak, associated to a direct band gap transition, that blueshifts upon uniaxial tension, indicating a tension-induced band gap increase. Interestingly, the spectra shift rate strongly depends on the direction along which the strain is





being applied. While strain along the *b*-axis yielded a spectral shift up to ~60-95 meV/%, strain applied along the *a*-axis only yielded a shift of ~0-15 meV/%. *Ab initio* calculations further verified the large anisotropic strain-tunable band-gap transitions observed in our experiments.

## 5. Materials and methods

*Sample fabrication:* Bulk ZrSe$_3$ crystals (from HQ Graphene) were exfoliated with Nitto SPV 224 tape and transferred to a Gel-Film (Gel-Pak WF ×4 6.0 mil) substrate. Then, the surface of the Gel-Film substrate is scanned under an optical microscope operated in transmission mode to identify the flakes. Once the desired flake is identified, it can be transferred onto a desired location of a target substrate by means of an all-dry deterministic placement method.[43–45]

*Strain dependent differential reflectance spectroscopy:* the disc-shape flexible substrate containing the desired flake on its center is loaded into the three-points bending system and the whole system is mounted under the objective of an optical microscope (Motic BA310 Met-T) system supplemented with a homebuilt micro-reflectance module based on a fiber-coupled CCD spectrometer (CCS200/M, Thorlabs) as previously reported.[48] The assigned flake is centered with respect to the central pivot using microscope inspection. Importantly, differential reflectance was used instead of photoluminescence, commonly employed to monitor the effect of strain on the optical properties of 2D semiconductors, given that photoluminescence measurements lead to a laser-induced burning of thin ZrSe$_3$ flakes.

*Ab initio calculations:* The electronic band structure and the optical reflectance have been calculated through the GW approximation starting from DFT calculations. We have initially relaxed the atomic structure, until the residual force between the atoms was below $4.0 \cdot 10^{-4}$ eV/Å. The lattice parameters, $a$=5.44 Å, $b$=3.73 Å and $c$= 9.45 Å with an angle between $a$ and $c$ of 97.6º were determined. All calculations are performed with scalar relativistic GGA-PBE





pseudopotentials in the optimised norm-conserving Vanderbilt set.[53,54] We used the Quantum Espresso suite for the calculations.[55,56] An energy cut-off of $10^3$ eV and a uniform mesh of k-points of 6×6×6 in the Brillouin zone were set. The GW calculations were performed using the Yambo code [57,58] starting from the Quantum Espresso output. The optical spectra have been obtained by solving the Bethe-Salpeter equation starting from the GW calculation in the plasmon-pole approximation. This calculation provides the real and imaginary part of the dielectric function of the material. From there, we have evaluated the reflectance considering that the light is perpendicular to the sample. The line width of the spectra is therefore obtained from the solution of that equation without any further approximation. In particular, the method is able to evaluate the electronic self-energy thus providing the line width. After modifying the unit cell parameters to take into account the uniaxial strain, we have optimised the atomic position till the residual force was below $4.0 \cdot 10^{-4}$ eV/Å.

*Scanning Transmission Electron Microscopy (STEM)*: For STEM characterization, ZrSe₃ flakes were mechanically transferred onto a holey $Si_3N_4$ membrane and characterized using an aberration-corrected JEOL JEM-ARM 200cF electron microscope equipped with a cold field emission gun and operated at 80 kV.

**Supporting Information**

Supporting Information is available from the Wiley Online Library or from the author. Supplementary Information includes: Atomic force microscopy characterization of the ZrSe3 flake measured in the main text (sample 1). Optical images of a ZrSe3 flake before and after Raman and photoluminescence testing. Baseline subtraction in the differential reflectance spectra. Differential reflectance spectra using non-polarized incident light. Effect of uniaxial compression on the reflectance spectra and strain tunable B exciton. Four-points bending configuration to apply compressive strain. Reproducibility test. Angle-resolved micro-





reflectance spectra of a ZrSe3 sample (Sample 8), under different uniaxial strain from 0%-0.65%, after 4 months of air exposure. Comparison between the strain tunable exciton energy in Sample 8 right after sample fabrication and after 4 months of storage in air. Optical images of Sample 11 taken at different days after its fabrication. Comparison between the strain tunable exciton energy (for strains parallel and perpendicular to b-axis) in Sample 11 acquired at different days after sample fabrication. Study of the anisotropic strain tunable reflectance spectra on samples 2, 3, 4, 5 and 6. Study of the maximum achievable strain before rupture, samples 7, 8 and 9. Study of the reproducibility of the strain tunable reflectance spectra, sample 10. Relationship between the optical contrast, extracted from the red channel of the optical images of the ZrSe3 flakes, with the thickness measured with AFM. The density of states (DOS) versus the electron energy for the monolayer, 5-layer and bulk ZrSe3. The band structure of ZrSe3 split into the atomic contributions.


### Acknowledgements

This project was funded from the European Research Council (ERC) under the European Union's Horizon 2020 research and innovation program (grant agreement n° 755655, ERC-StG 2017 project 2D-TOPSENSE). R.F. acknowledges the support from the Spanish Ministry of Economy, Industry and Competitiveness through a Juan de la Cierva-formación fellowship 2017 FJCI2017-32919. R.D'A acknowledges the financial support of the Grupos Consolidados del Gobierno Basco (grant IT1249-19) and the MINECO (grant G17/A01). G.S.-S. acknowledges financial support from Spanish MICIU RTI2018-099054-J-I00 and MICINN IJC2018-038164-I. Electron microscopy observations were carried out at the Centro Nacional de Microscopia Electronica, CNME-UCM. R.D'A also acknowledges useful discussions with D. Varsano. H. L. acknowledges the grant from China Scholarship Council (CSC) under No.






201907040070. The authors extend their sincere appreciation to the Distinguished Scientist

Fellowship Program (DSFP) at King Saud University for funding of this work.

Zirconium triselenide is a two-dimensional semiconductor material with an interesting in-plane anisotropic crystal structure. Here we demonstrate that the excitons of this material are very sensitive to uniaxial strains and, interestingly, their tunability strongly depends on the direction along which the uniaxial strain is applied.


*Hao Li, Gabriel Sanchez-Santolino, Sergio Puebla, Riccardo Frisenda, Abdullah M. Al-Enizi, Ayman Nafady, Roberto D'Agosta*, Andres Castellanos-Gomez**


**Strongly anisotropic strain-tunability of excitons in exfoliated ZrSe₃**

ToC figure

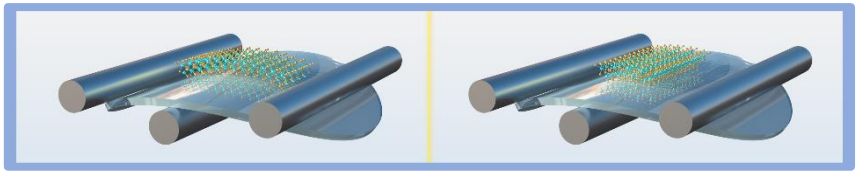





# Supporting Information

## Strongly anisotropic strain-tunability of excitons in exfoliated ZrSe₃


*Hao Li, Gabriel Sanchez-Santolino, Sergio Puebla, Riccardo Frisenda, Abdullah M. Al-Enizi, Ayman Nafady, Roberto D'Agosta\*, Andres Castellanos-Gomez\**


**Atomic force microscopy characterization of the ZrSe₃ flake measured in the main text (sample 1)**
**Optical images of a ZrSe₃ flake before and after Raman and photoluminescence testing**
**Baseline subtraction in the differential reflectance spectra**
**Differential reflectance spectra using non-polarized incident light**
**Effect of uniaxial compression on the reflectance spectra and strain tunable B exciton**
**Four-points bending configuration to apply compressive strain**
**Reproducibility test**
**Angle-resolved micro-reflectance spectra of a ZrSe₃ sample (Sample 8), under different uniaxial strain from 0%-0.65%, after 4 months of air exposure**
**Comparison between the strain tunable exciton energy in Sample 8 right after sample fabrication and after 4 months of storage in air**
**Optical images of Sample 11 taken at different days after its fabrication**
**Comparison between the strain tunable exciton energy (for strains parallel and perpendicular to b-axis) in Sample 11 acquired at different days after sample fabrication**
**Study of the anisotropic strain tunable reflectance spectra on samples 2, 3, 4, 5 and 6**
**Study of the maximum achievable strain before rupture, samples 7, 8 and 9**
**Study of the reproducibility of the strain tunable reflectance spectra, sample 10**
**Relationship between the optical contrast, extracted from the red channel of the optical images of the ZrSe₃ flakes, with the thickness measured with AFM**
**The density of states (DOS) versus the electron energy for the monolayer, 5-layer and bulk ZrSe₃**
**The band structure of ZrSe3 split into the atomic contributions.**





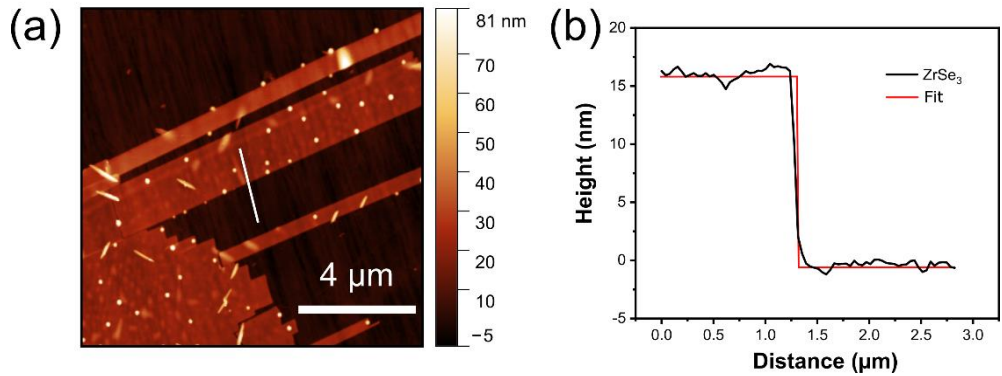

**Figure S1. Atomic force microscopy characterization of the ZrSe₃ flake measured in the main text (sample 1).** (a) AFM topography image. (b) Line profile measured along the white line in (a) to show the thickness of the flake.

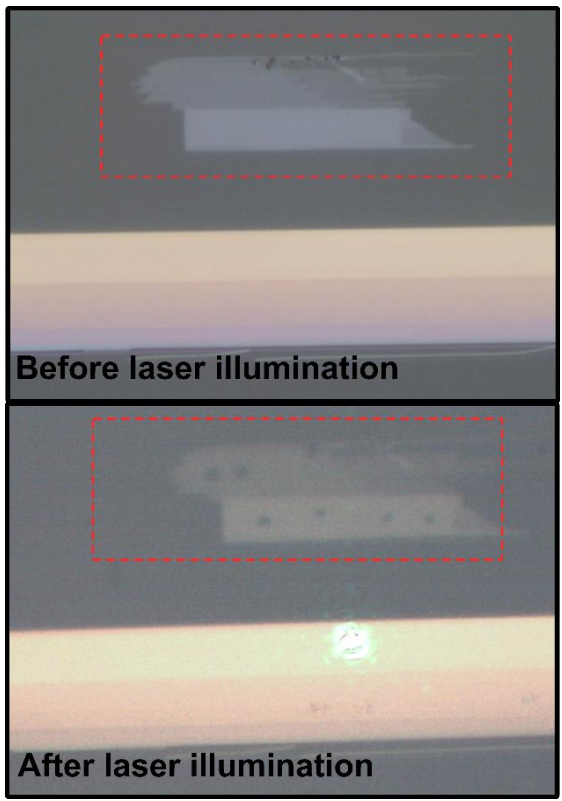

**Figure S2. Optical images of a ZrSe₃ flake before and after Raman and photoluminescence testing.** We could not observe any Raman or PL spectra because of the laser-induced damage. Wavelength: 532 nm, objective: 50×. Acquisition time: 20 s, power: 0.129 mW, spot size: 2 μm, power density: $4.1 \cdot 10^{-5}$ W/μm$^2$.





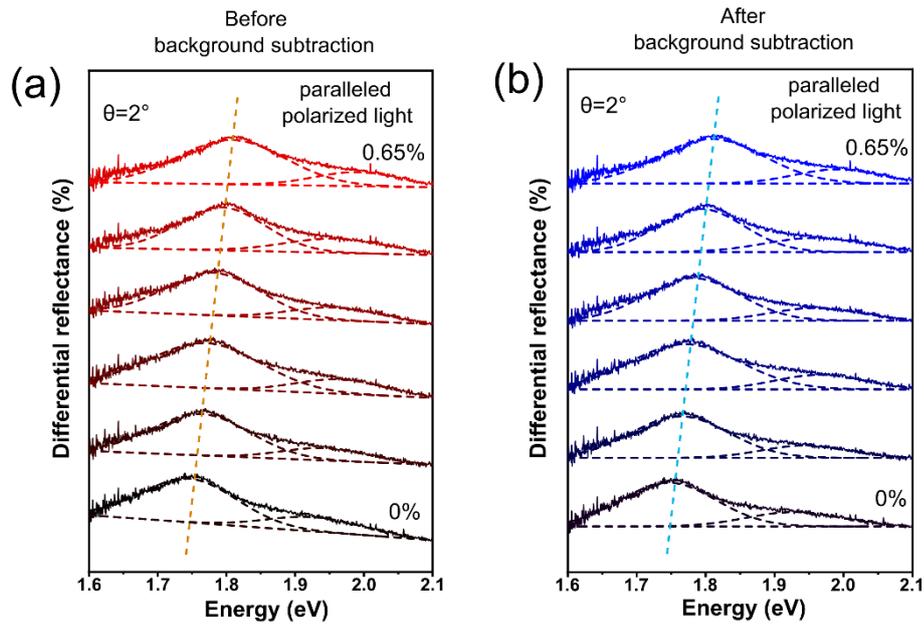

**Figure S3. Baseline subtraction in the differential reflectance spectra.** (a) Differential reflectance spectra acquired (sample 1) for different uniaxial strains applied along the *b*-axis before subtracting the background. (b) Same spectra after background subtraction.





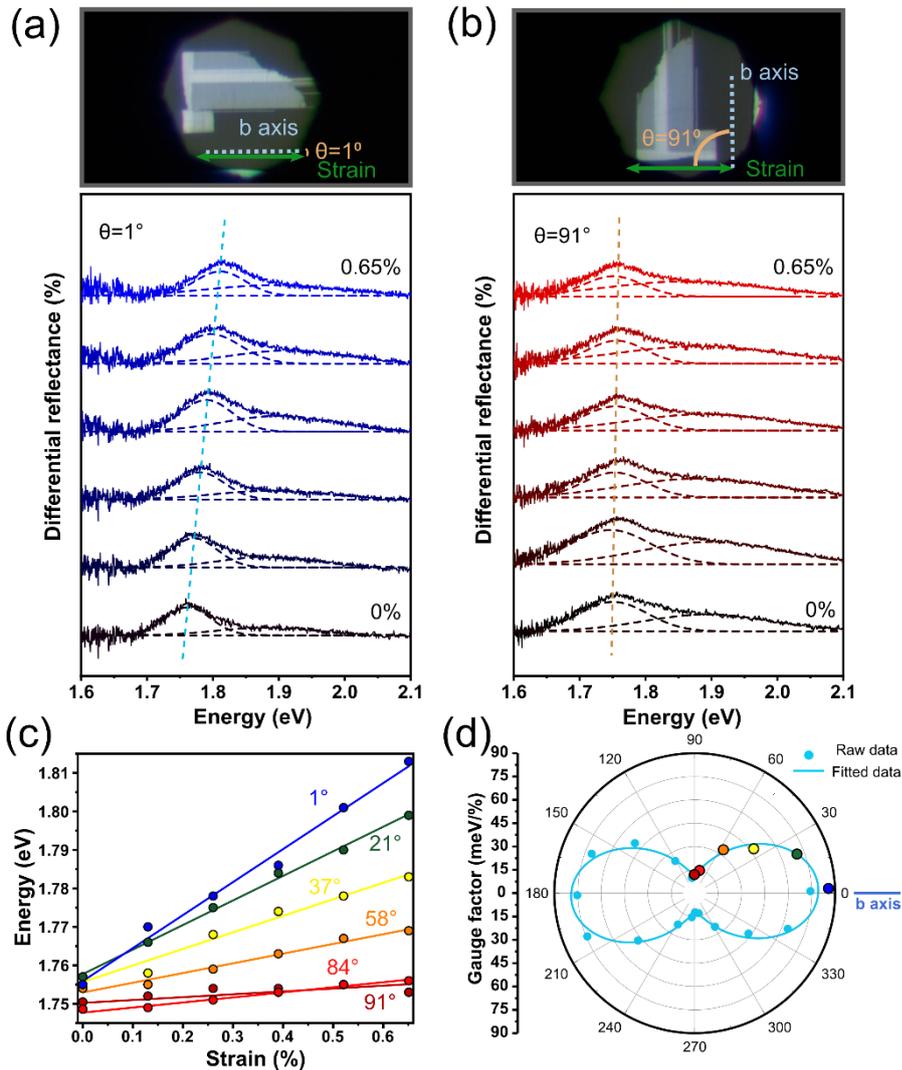

**Figure S4. Differential reflectance spectra using non-polarized incident light.** Same figure as Figure 4 of the main text (sample 1) but using non-polarized light instead of linearly polarized light along the *b*-axis. The exciton peaks are less pronounced than using linearly polarized light but the strain tunability is similar in magnitude and trend. (a, b) The micro-reflectance spectra acquired when the uniaxial strain direction is parallel to the *b*-axis and the *a*-axis of the ZrSe₃ flake, respectively. (c) The energy of the excitonic peak as a function of the applied uniaxial strain under different orientation angles (1º, 21º, 37º, 58º, 84º and 91º). A linear fit is utilized for extracting the gauge factor. (d) The experimental and fitted gauge factor change of the excitonic peak of ZrSe₃ in polar coordinates.





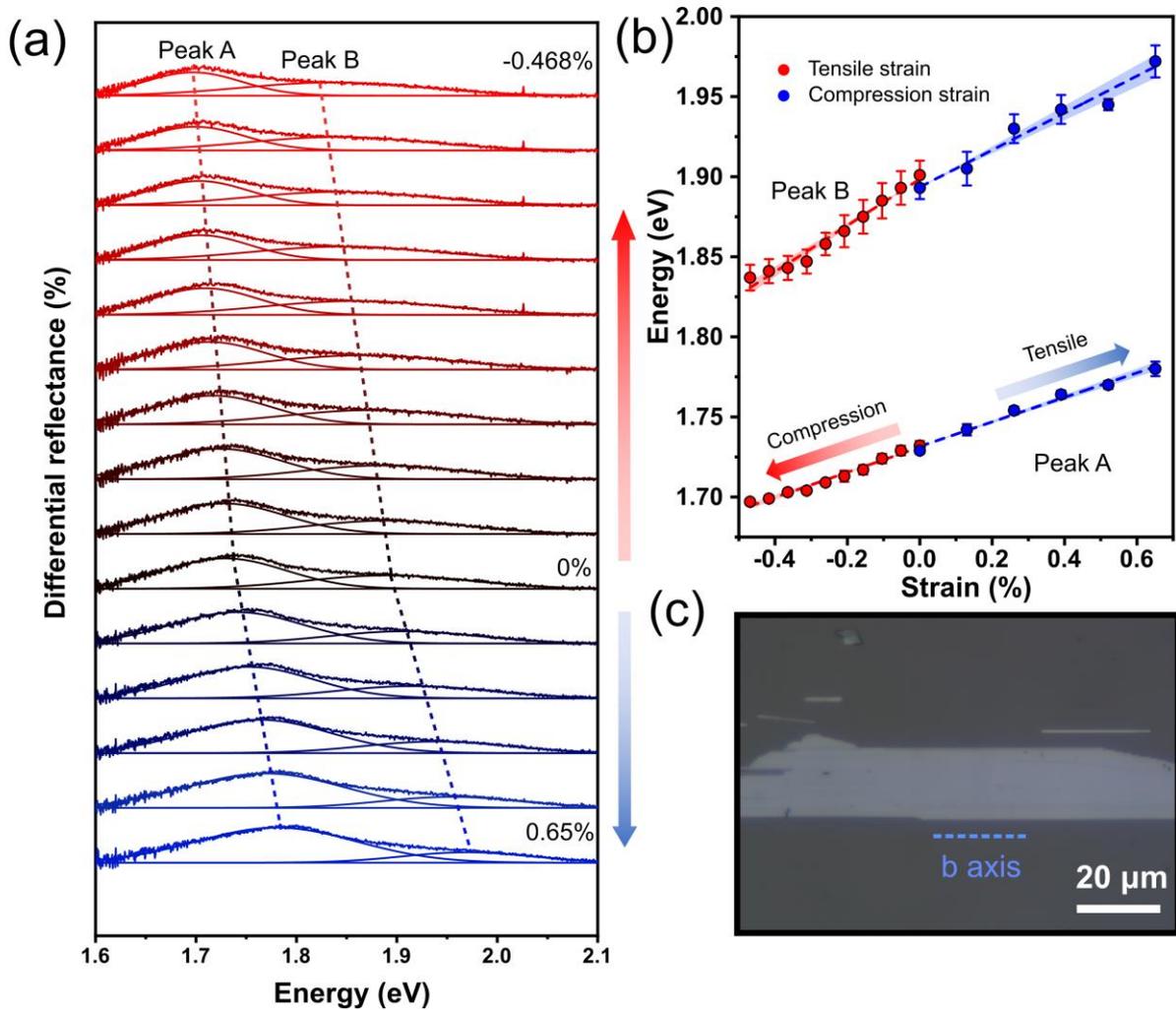

**Figure S5. Effect of uniaxial compression on the reflectance spectra and strain tunable B exciton.** (a) Differential reflectance spectra acquired on Sample 10 for different uniaxial strains along b ranging from -0.468% (compression) to +0.65% (tension). (b) Energy of the A and B exciton peaks as a function of strain. (c) Optical microscopy image of the sample 10. A four-points bending setup has been used for this measurement (see Scheme S1).

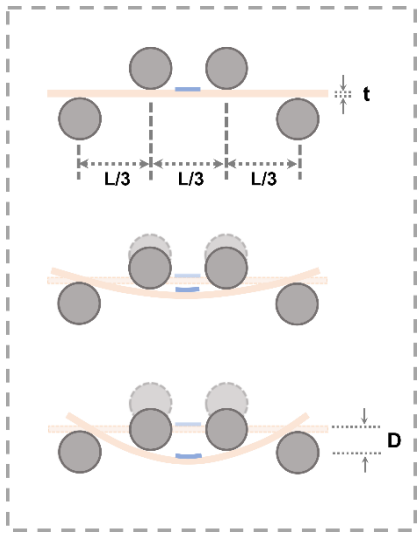





**Scheme S1. Four-points bending configuration to apply compressive strain.** The four-points bending setup calculation is based on reference [S1]: $\varepsilon = 27 \ast D \ast t / (5 \cdot L^2)$. D represents the deflection of the plate. t represents the thickness of the plate, and L is the distance between the two outer cylinders (36 mm).





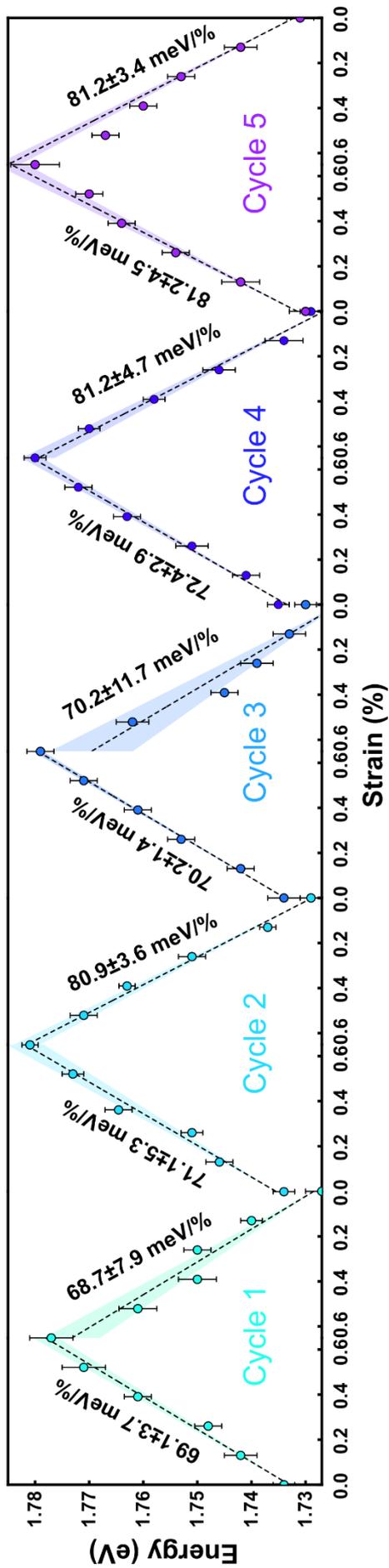

**Figure S6. Reproducibility test.** A exciton energy as a function of uniaxial tensile strain for 5 different cycles of strain loading/unloading. The measurements have been carried out in sample 10.





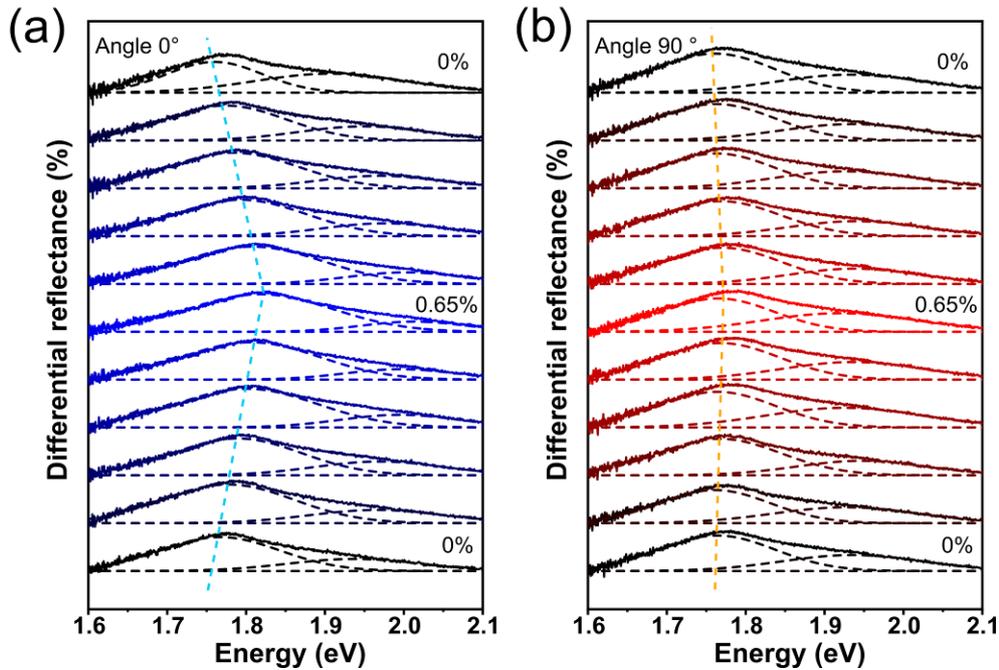

**Figure S7. Angle-resolved micro-reflectance spectra of a ZrSe₃ sample (Sample 8), under different uniaxial strain from 0%-0.65%, after 4 months of air exposure.** (a, b) Micro-reflectance spectra acquired when the uniaxial strain direction is parallel to the *b*-axis and the *a*-axis of the ZrSe₃ flake, respectively.

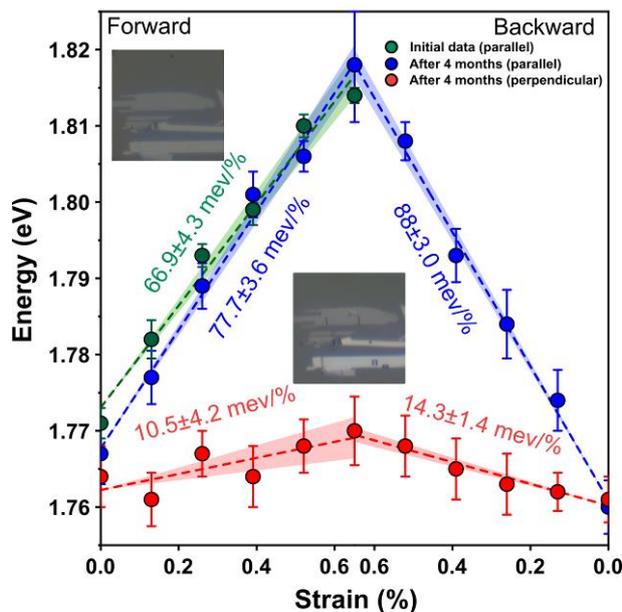

**Figure S8. Comparison between the strain tunable exciton energy in Sample 8 right after sample fabrication and after 4 months of storage in air.** (Inset, up) Optical image of the flake right after exfoliation and transfer. (Inset, lower) Optical image of the flake after 4 months in air.





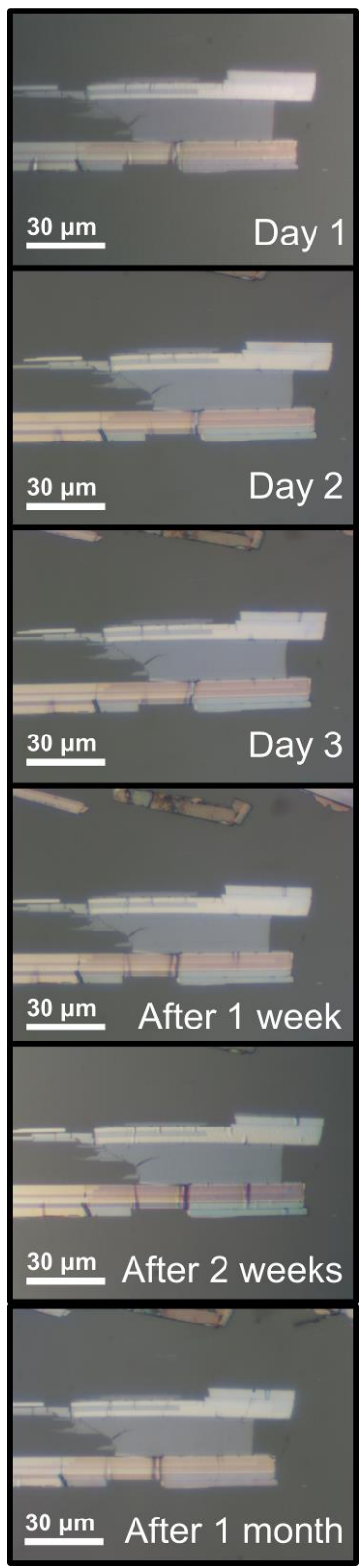

**Figure S9. Optical images of Sample 11 taken at different days after its fabrication.**









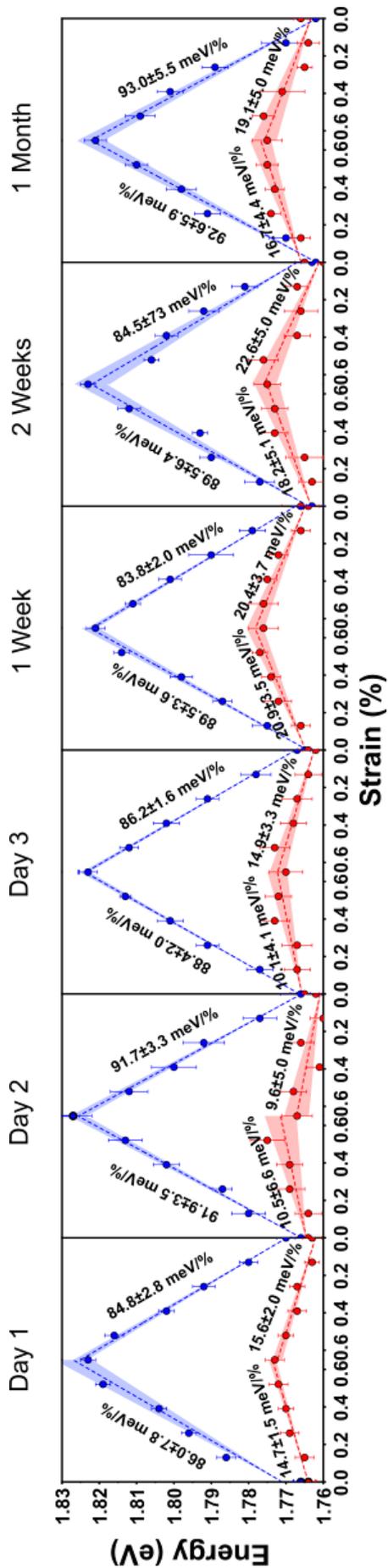

**Figure S10. Comparison between the strain tunable exciton energy (for strains parallel and perpendicular to b-axis) in Sample 11 acquired at different days after sample fabrication**.





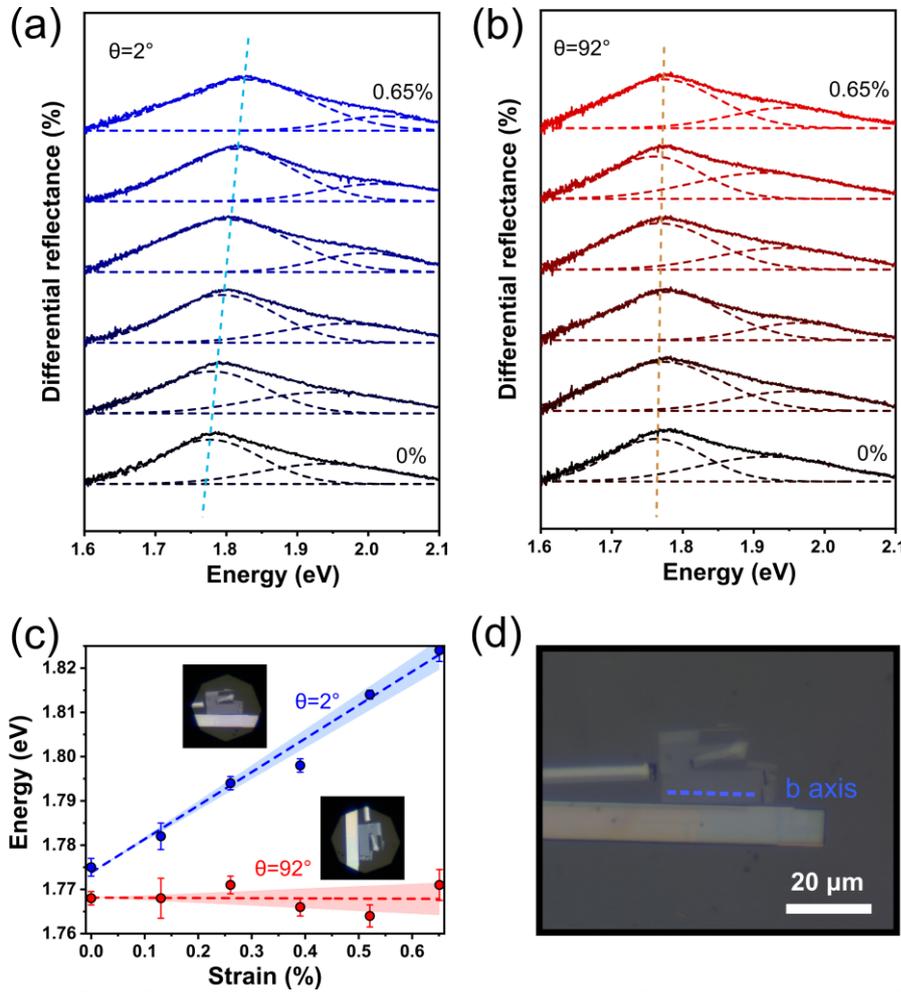

**Figure S11. Study of the anisotropic strain tunable reflectance spectra on Sample 2.** (a, b) The micro-reflectance spectra acquired when the uniaxial strain direction is parallel to the *b*-axis and the *a*-axis of the ZrSe₃ flake, respectively. (c) The energy of the A excitonic peak as a function of the applied uniaxial strain for the parallel and perpendicular to *b*-axis configurations. A linear fit is utilized for extracting the gauge factor. (d) Optical microscopy image of the sample 2.





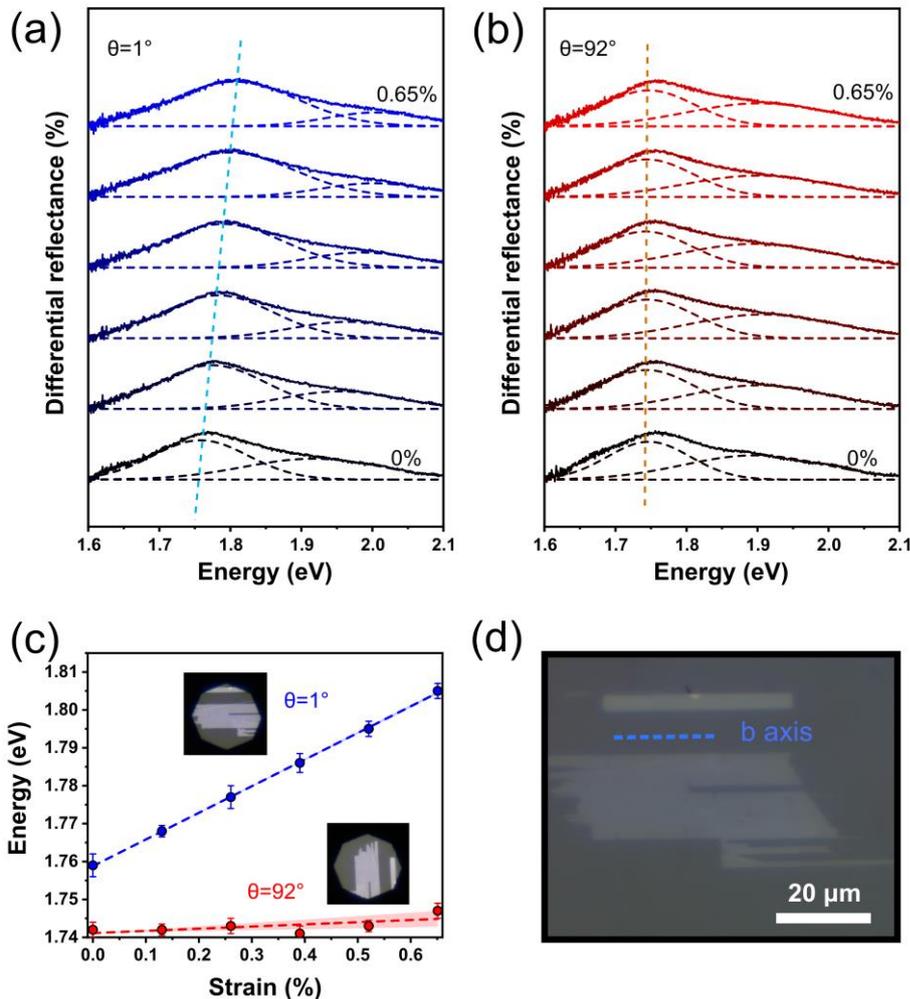

**Figure S12. Study of the anisotropic strain tunable reflectance spectra on Sample 3.** (a, b) The micro-reflectance spectra acquired when the uniaxial strain direction is parallel to the *b*-axis and the *a*-axis of the ZrSe₃ flake, respectively. (c) The energy of the A excitonic peak as a function of the applied uniaxial strain for the parallel and perpendicular to *b*-axis configurations. A linear fit is utilized for extracting the gauge factor. (d) Optical microscopy image of the sample 3.





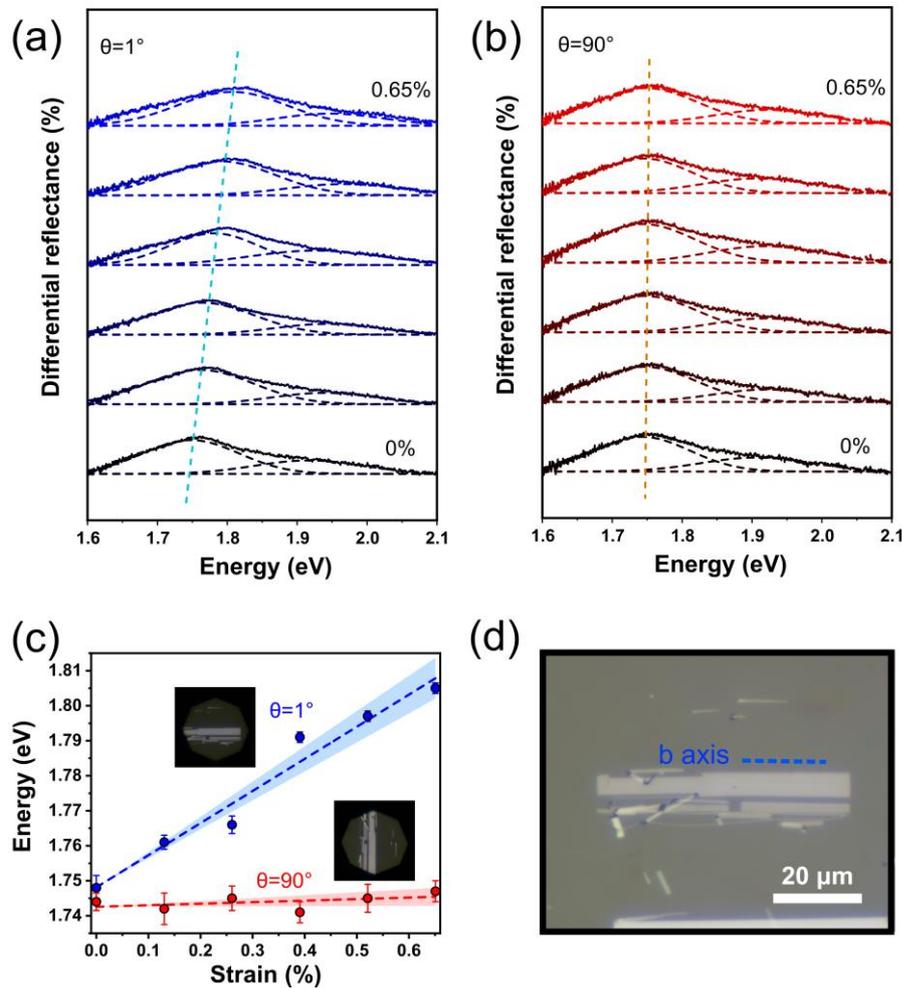

**Figure S13. Study of the anisotropic strain tunable reflectance spectra on Sample 4.** (a, b) The micro-reflectance spectra acquired when the uniaxial strain direction is parallel to the *b*-axis and the *a*-axis of the ZrSe$_3$ flake, respectively. (c) The energy of the A excitonic peak as a function of the applied uniaxial strain for the parallel and perpendicular to *b*-axis configurations. A linear fit is utilized for extracting the gauge factor.  (d) Optical microscopy image of the sample 4.





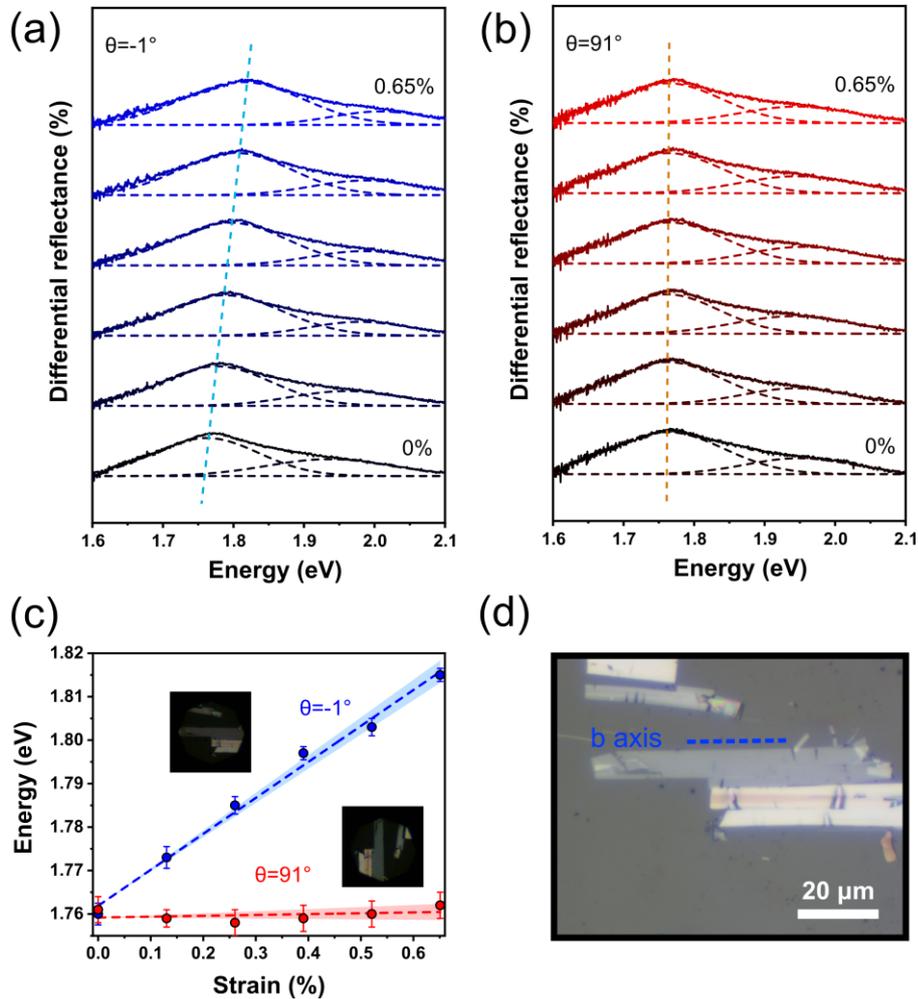

**Figure S14. Study of the anisotropic strain tunable reflectance spectra on Sample 5.** (a, b) The micro-reflectance spectra acquired when the uniaxial strain direction is parallel to the *b*-axis and the *a*-axis of the ZrSe$_3$ flake, respectively. (c) The energy of the A excitonic peak as a function of the applied uniaxial strain for the parallel and perpendicular to *b*-axis configurations. A linear fit is utilized for extracting the gauge factor. (d) Optical microscopy image of the sample 5.





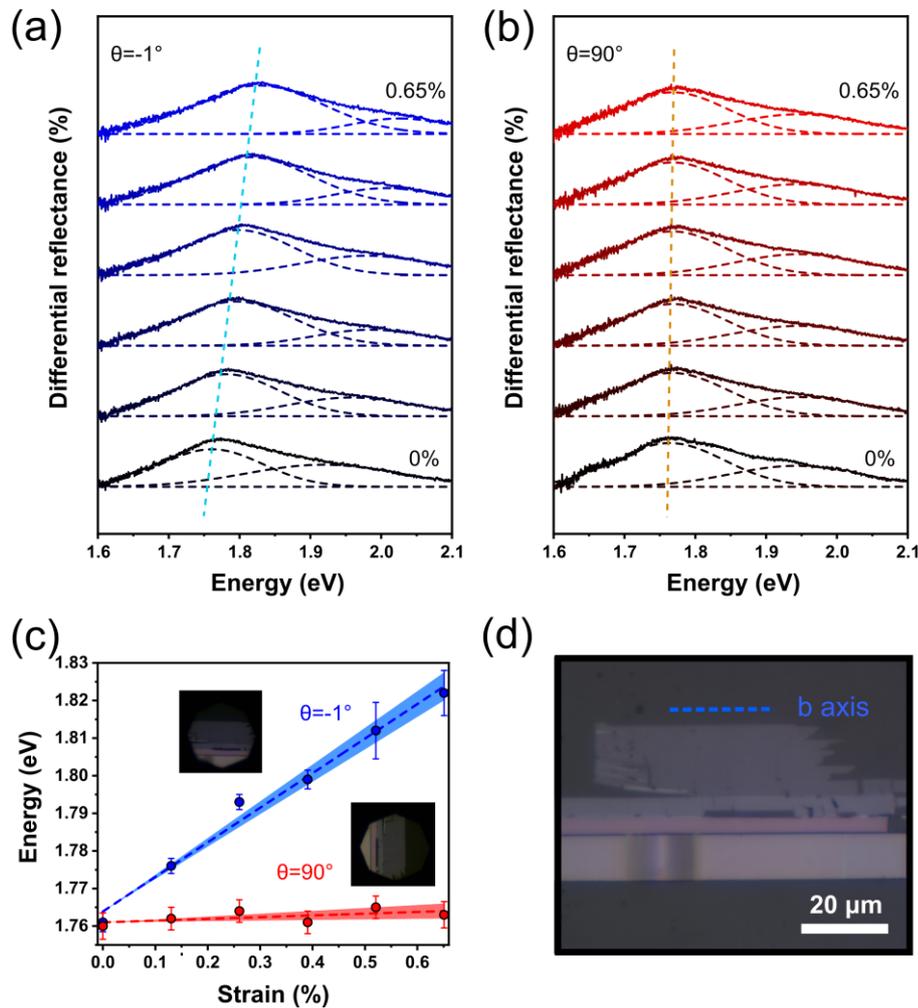

**Figure S15. Study of the anisotropic strain tunable reflectance spectra on Sample 6.** (a, b) The micro-reflectance spectra acquired when the uniaxial strain direction is parallel to the *b*-axis and the *a*-axis of the ZrSe₃ flake, respectively. (c) The energy of the A excitonic peak as a function of the applied uniaxial strain for the parallel and perpendicular to *b*-axis configurations. A linear fit is utilized for extracting the gauge factor.  (d) Optical microscopy image of the sample 6.





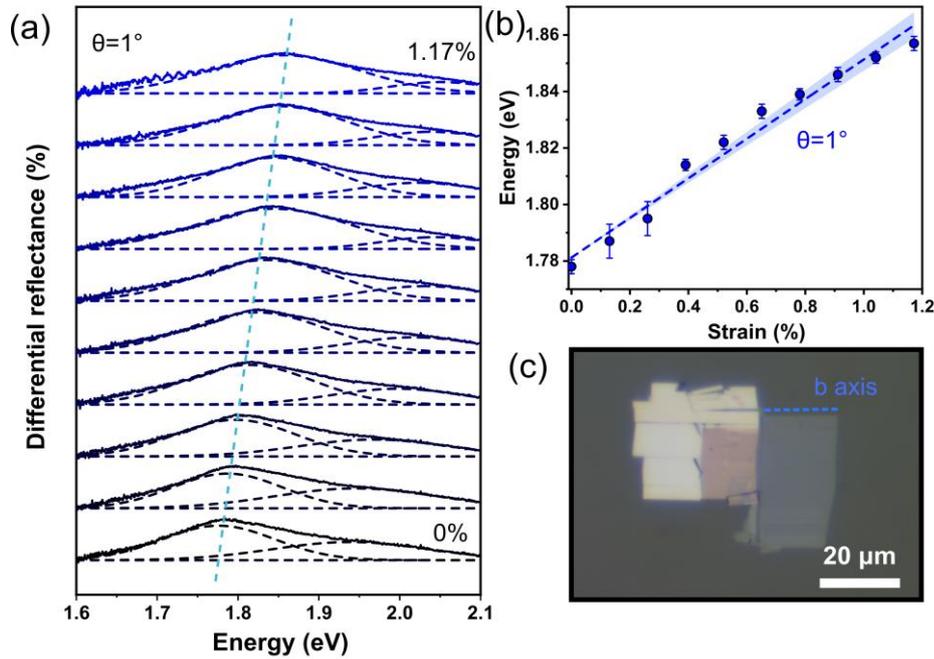

**Figure S16. Study of the maximum achievable strain before rupture, Sample 7.** (a) Micro-reflectance spectra acquired when the uniaxial strain direction is parallel to the *b*-axis of the ZrSe₃ flake. (b) The energy of the A excitonic peak as a function of the applied uniaxial strain for the parallel to *b*-axis configuration. (c) Optical microscopy image of the sample 7 before and after strain-induced rupture.

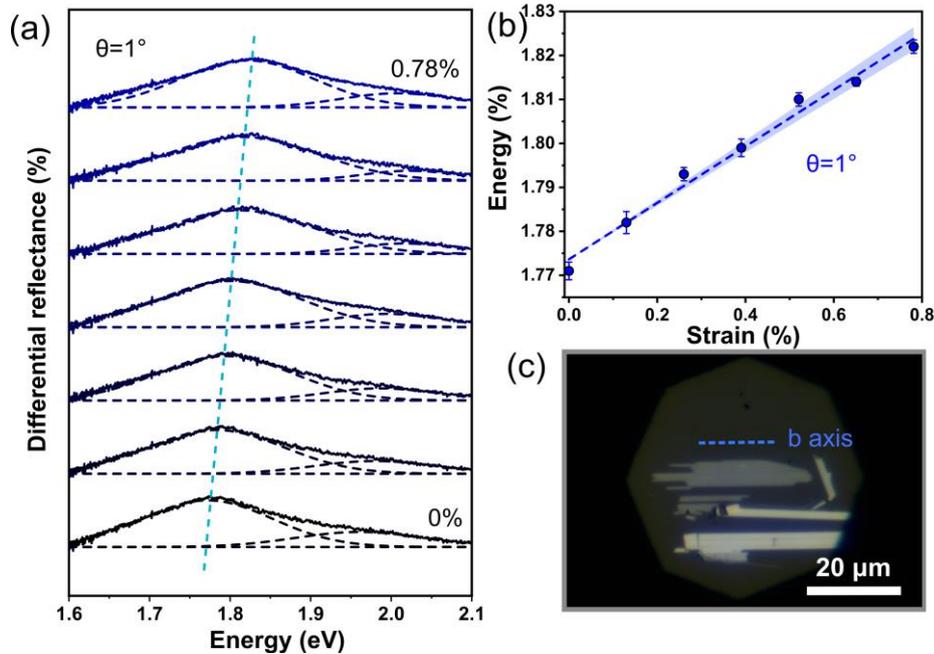

**Figure S17. Study of the maximum achievable strain before rupture, Sample 8.** (a) Micro-reflectance spectra acquired when the uniaxial strain direction is parallel to the *b*-axis of the ZrSe₃ flake. (b) The energy of the A excitonic peak as a function of the applied uniaxial strain for the parallel to *b*-axis configuration. (c) Optical microscopy image of the sample 8.





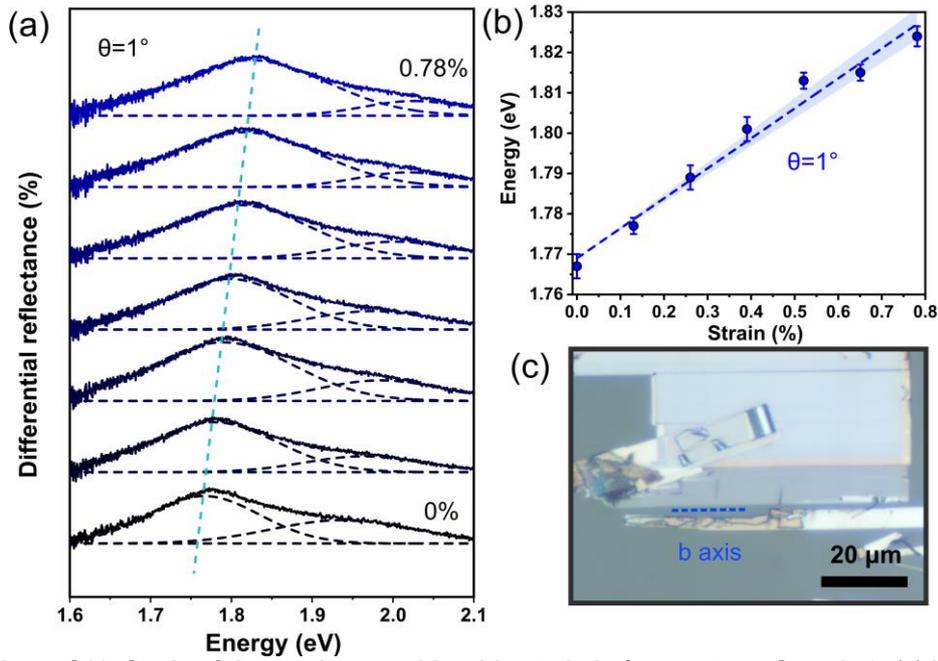

**Figure S18. Study of the maximum achievable strain before rupture, Sample 9.** (a) Micro-reflectance spectra acquired when the uniaxial strain direction is parallel to the *b*-axis of the ZrSe₃ flake. (b) The energy of the A excitonic peak as a function of the applied uniaxial strain for the parallel to *b*-axis configuration. (c) Optical microscopy image of the sample 9 before and after strain-induced rupture.





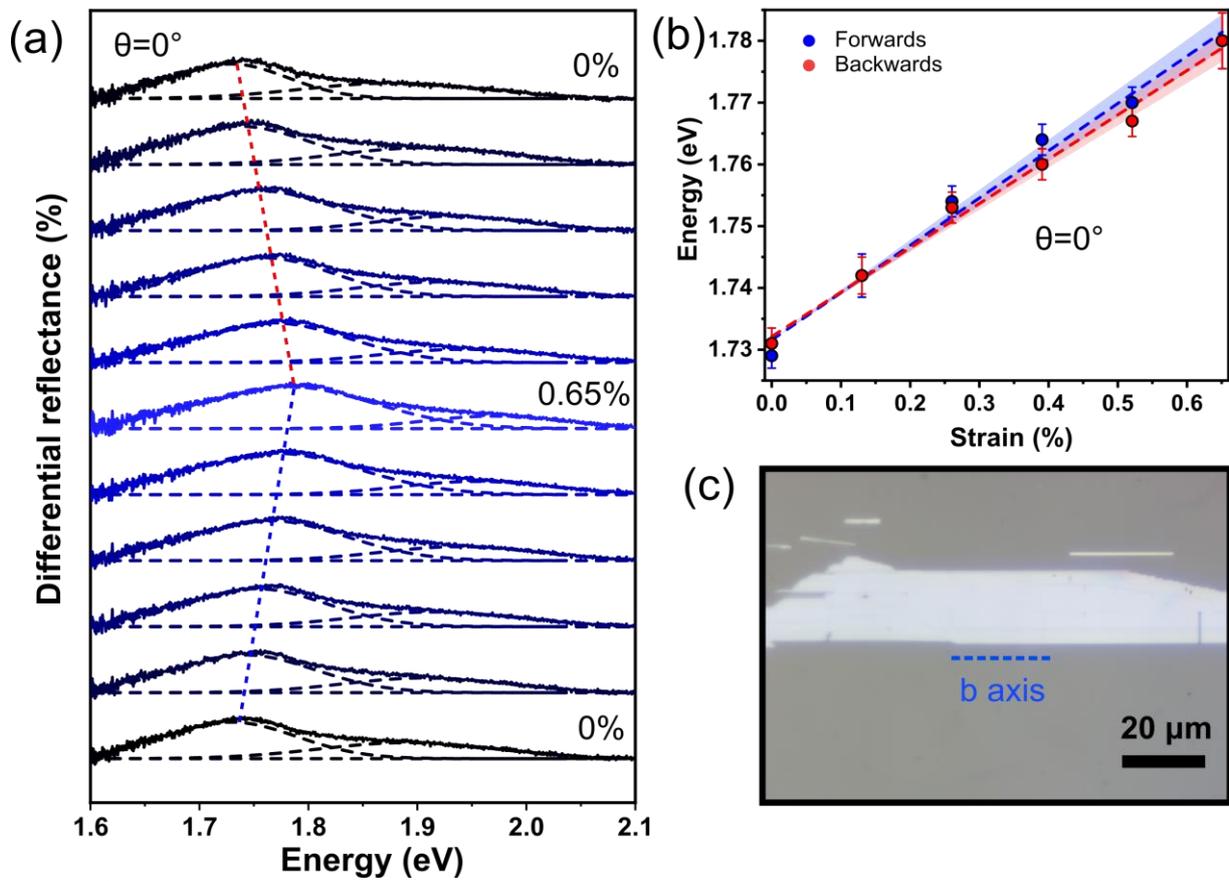

**Figure S19. Study of the reproducibility of the strain tunable reflectance spectra, Sample 10.** (a) Micro-reflectance spectra acquired when the uniaxial strain direction is parallel to the *b*-axis of the ZrSe₃ flake. (b) The energy of the A excitonic peak as a function of the applied uniaxial strain for the parallel to *b*-axis configuration. (c) Optical microscopy image of the sample 10.

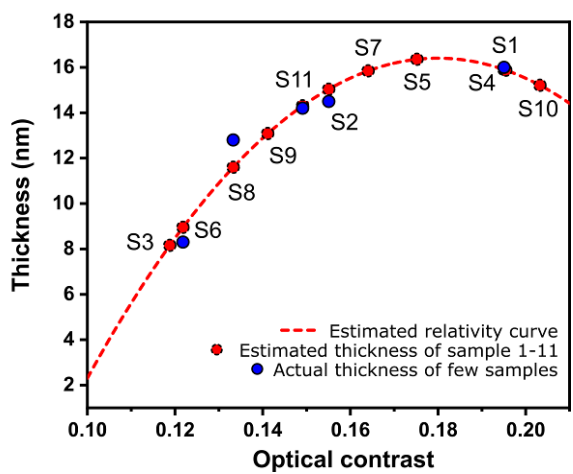

**Figure S20. Relationship between the optical contrast, extracted from the red channel of the optical images of the ZrSe₃ flakes, with the thickness measured with AFM (blue dots).** The experimental data has been fitted to a quadratic polynomial to estimate the thickness of the rest of the ZrSe₃ flakes from their optical contrast values.





## Density of states (DOS): monolayer to bulk transition

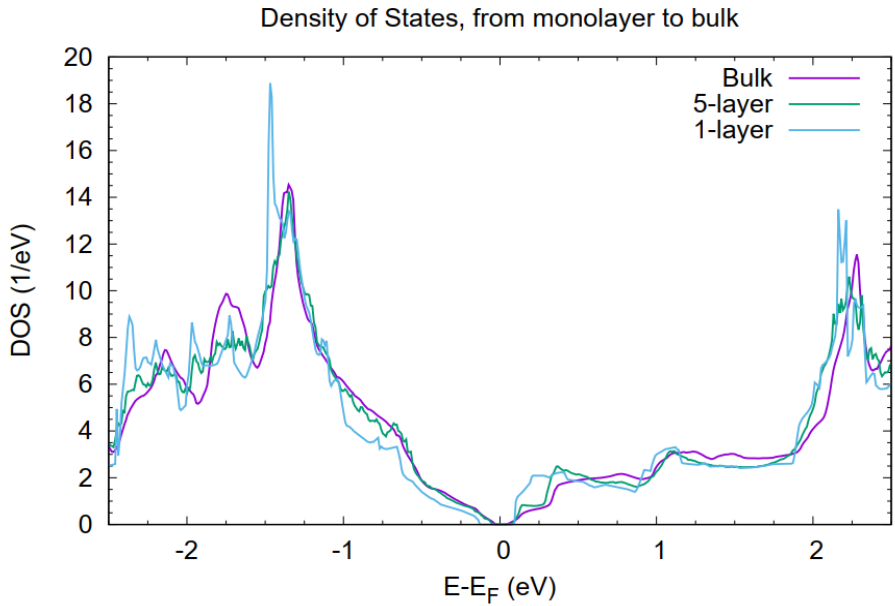

Density of States, from monolayer to bulk

**Figure S21.** The density of states (DOS) versus the electron energy for the monolayer, 5-layer and bulk ZrSe$_3$.

In Figure S21, we report the density of states as a function of the energy for the monolayer, 5-layer and bulk ZrSe$_3$. Although there remain certain differences dictated mostly by the periodic nature of the bulk, the 5-layer and bulk DOS show similar features especially in the window -1 to 1 eV around the Fermi energy, namely the region of interest for optical applications.

## Band structure analysis

To understand the anisotropic behaviour of the shift in the exciton peaks with applied strain, we have analysed the contribution to the band structure by the different atoms [S2]. Figure S22 reports the band structure projected on each of the atoms in the unit cell. The contribution is split on the Zr and two Se atoms. Indeed, not all the Se atoms are equivalent. Indeed, Se$_2$ here is the atom more internal and binds the two Zr atoms. Se$_1$ instead is one of the external atoms which binds with van der Walls forces the multilayer structure. From the figure, we see that the valence band at the $\Gamma$ point is dominated by the contribution of the Se$_2$ atom, while the Zr determines the conduction bands. Thus, we focus on those two atoms. (Notice that there are 4 Se atoms equivalent to Se$_1$, 2 Se atoms equivalent to Se$_2$, and two Zr atoms.)





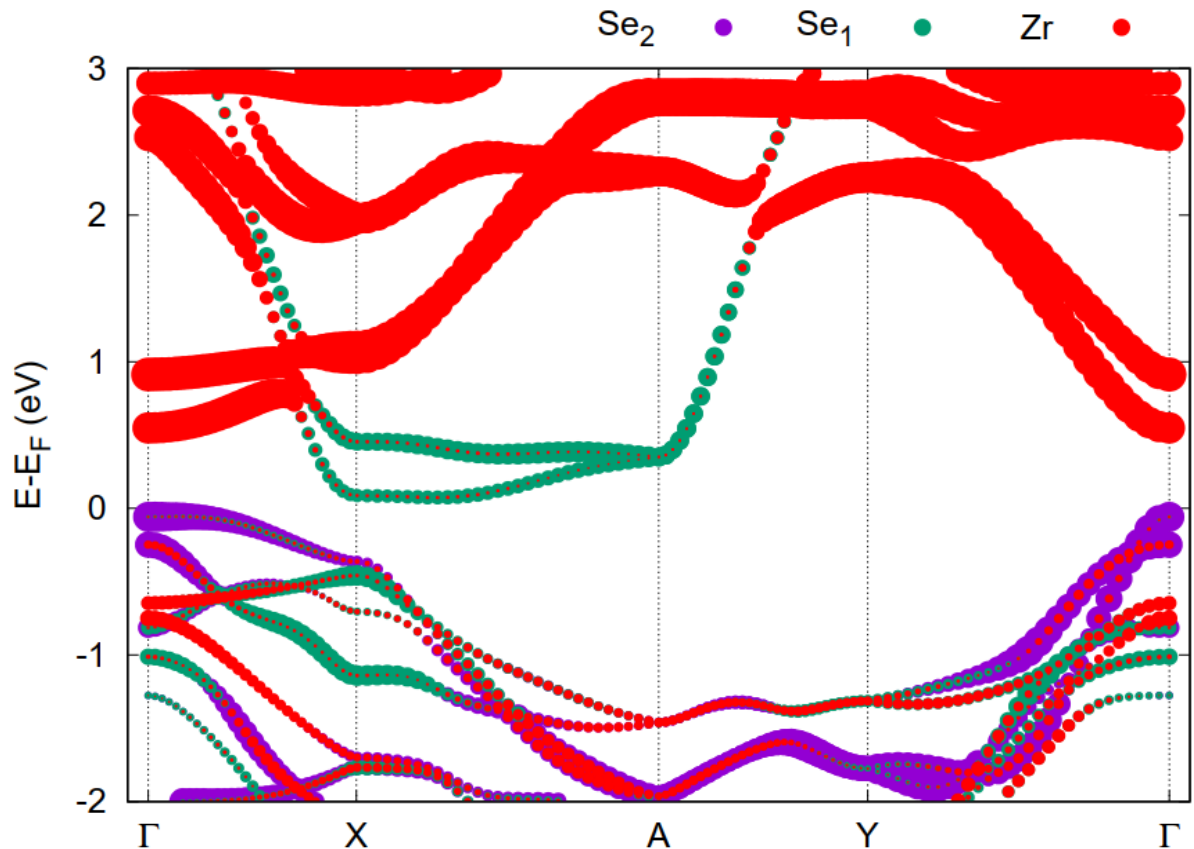

**Figure S22:** The band structure of ZrSe₃ split into the atomic contributions.

In Figure S23, we split the contribution of the Zr atoms in the contribution of its d states. Our analysis allows us to conclude that the conduction bands, around Γ, are mainly determined by the Zr $d_{xy}$ and $d_{yz}$ orbitals. The $d_{xz}$ orbital has a contribution only at large energy. This fact suggests that deforming the unit cell along y (or b) might have a more significant impact since it modifies the bonds containing the $d_{xy}$ and $d_{yz}$ orbitals. At the same time, a deformation along x (or a) only affects the $d_{xz}$ orbital. We complete this analysis by looking at the Se₂ atoms and splitting the contribution of the atomic states (see Figure S24). We can see from there that the $p_z$ orbital dominates the valence band maximum around the Γ point, while the $p_y$ contributes the most to the second to last valence state. This observation points to the direction of a certain insensitiveness of the valence band maximum to strain applied to both a and b direction (parallel to x and y in our case) as pointed out by our band structure calculations.





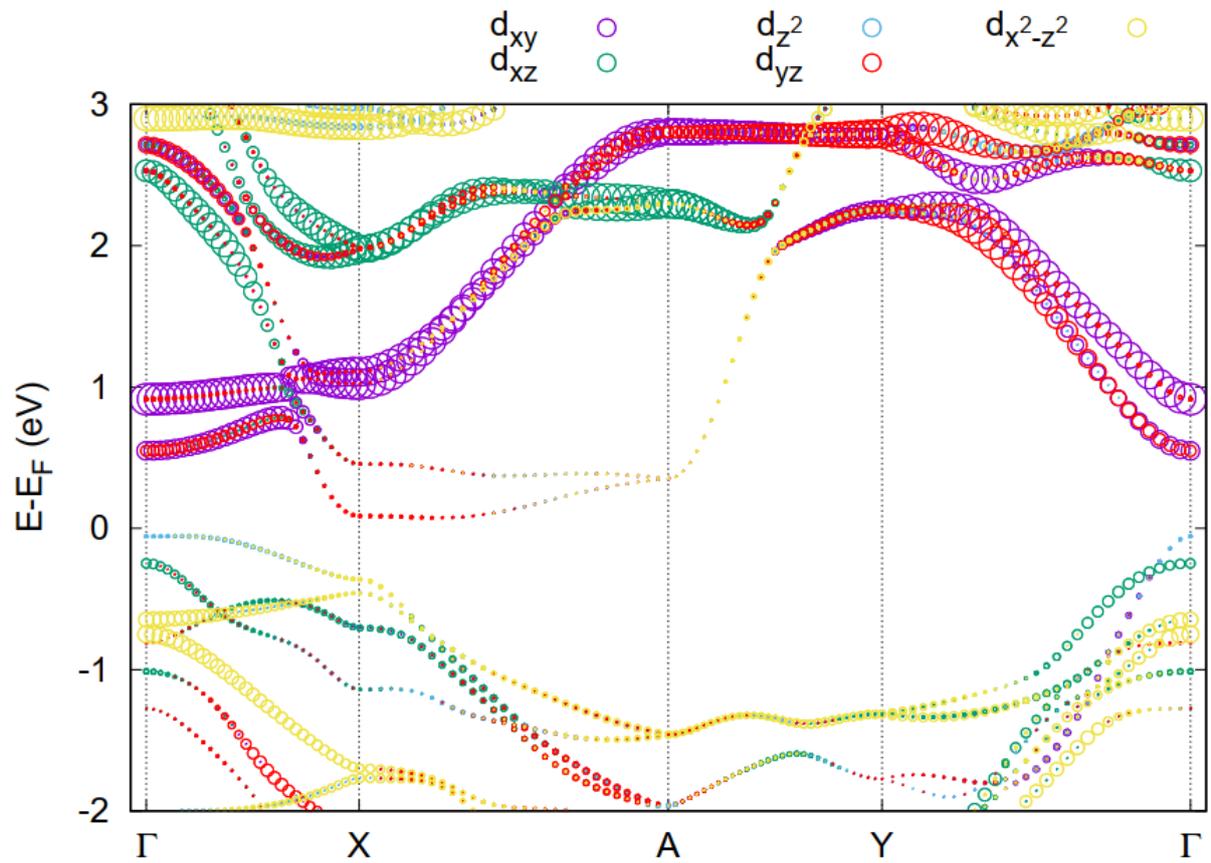

**Figure S23.** We thus project the bands into the contribution of the d orbitals of the Zr atoms. The s and p orbital do not contribute to this energy range.





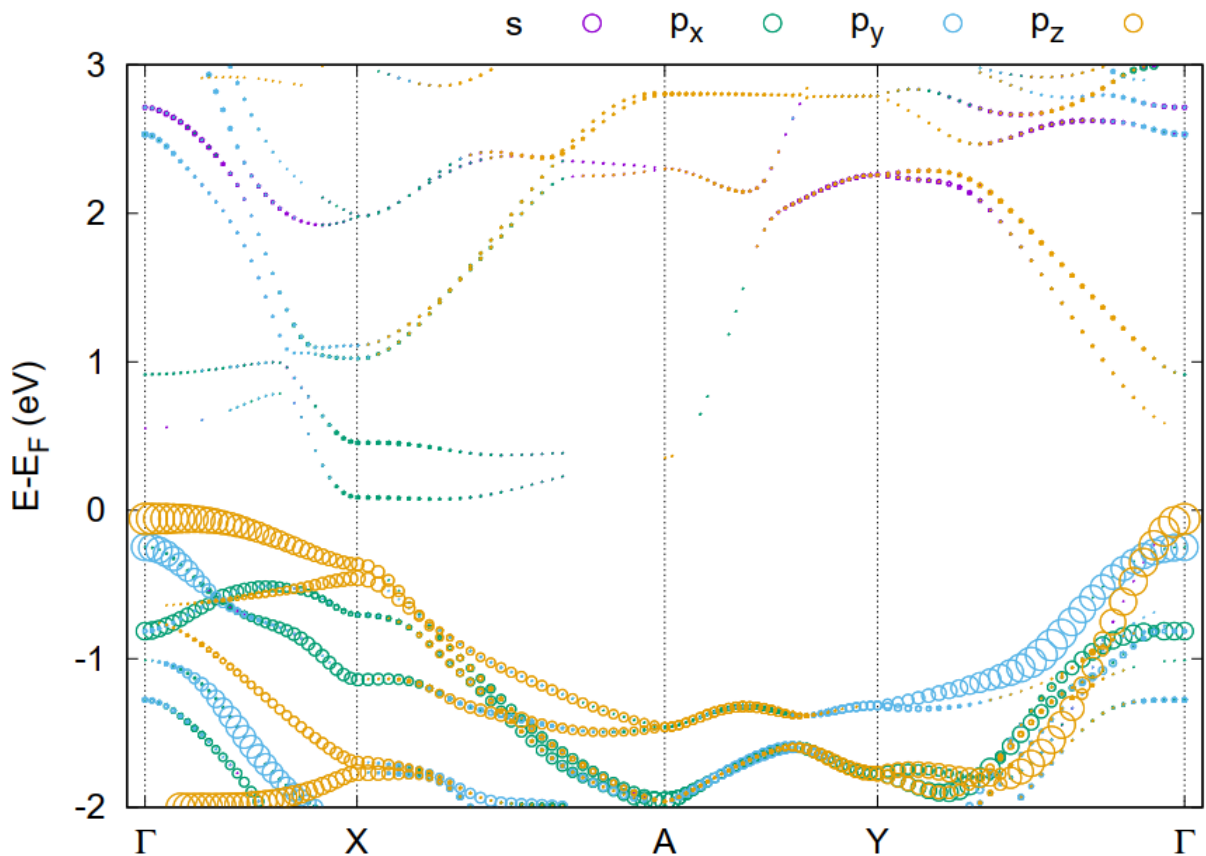

**Figure S24.** Band structure of ZrSe₃ projected along the s and p orbital of the Se₂ atoms.

Finally, in Figure S25, we plot the electronic density with different point of view: a) along the a-axis, b) along the b-axis. We can notice the anisotropic structure of the material. Electron density accumulate more between the Zr and Se atoms along the b axis rather than along a. This explains the different response of the system to the two uniaxial strains.

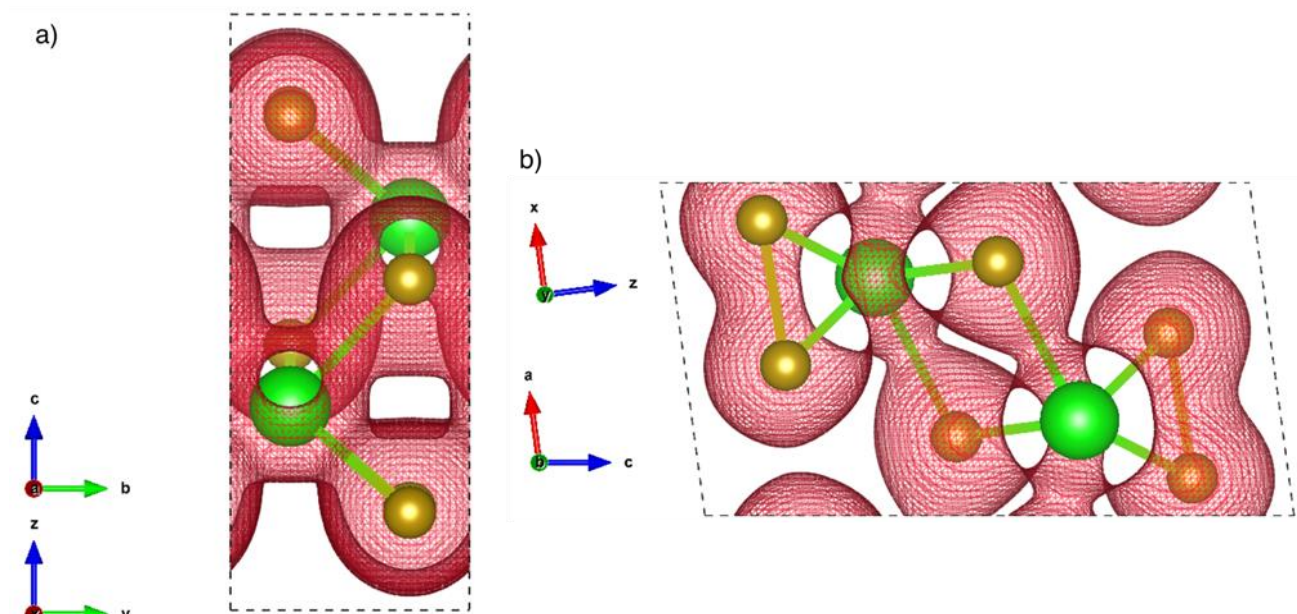

**Figure S25.** Electron density as projected along the a) a- and b) b-axis. The figure is obtained with VESTA [S3].





Supporting Information References